\documentclass[superscriptaddress,showpacs,twocolumn,aps,pre,10pt]{revtex4-1}

\usepackage{graphicx,dcolumn,bm,bbm,amsmath,amssymb,color}
\usepackage[squaren]{SIunits}
\usepackage[english]{babel}

\begin{document}

\title{Non-Gaussian statistics for the motion of self-propelled Janus
particles: experiment versus theory}

\author{Xu Zheng}
\affiliation{State Key Laboratory of Nonlinear Mechanics, Institute of Mechanics, CAS, Beijing 100190, People's Republic of China}

\author{\surname{Borge} ten Hagen}
\email{bhagen@thphy.uni-duesseldorf.de}
\affiliation{Institut f{\"u}r Theoretische Physik II: Weiche Materie, Heinrich-Heine-Universit{\"a}t D{\"u}sseldorf, D-40225 D{\"u}sseldorf, Germany}

\author{Andreas Kaiser}
\affiliation{Institut f{\"u}r Theoretische Physik II: Weiche Materie, Heinrich-Heine-Universit{\"a}t D{\"u}sseldorf, D-40225 D{\"u}sseldorf, Germany}

\author{Meiling~Wu}
\affiliation{Xi`an University of Architecture and Technology, Xi`an, 710055, People's Republic of China}

\author{Haihang Cui}
\affiliation{Xi`an University of Architecture and Technology, Xi`an, 710055, People's Republic of China}

\author{Zhanhua Silber-Li}
\email{lili@imech.ac.cn}
\affiliation{State Key Laboratory of Nonlinear Mechanics, Institute of Mechanics, CAS, Beijing 100190, People's Republic of China}

\author{Hartmut L{\"o}wen}
\affiliation{Institut f{\"u}r Theoretische Physik II: Weiche Materie, Heinrich-Heine-Universit{\"a}t D{\"u}sseldorf, D-40225 D{\"u}sseldorf, Germany}

\date{\today}

\begin{abstract}
Spherical Janus particles are one of the most prominent examples for active Brownian objects. Here, we study the diffusiophoretic motion of such microswimmers in experiment and in theory. Three stages are found: simple Brownian motion at short times, super-diffusion at intermediate times, and finally diffusive behavior again at long times.
These three regimes observed in the experiments are compared with a theoretical model for the Langevin dynamics of self-propelled particles with coupled translational and rotational motion. Besides the mean square displacement also higher displacement moments are addressed.
In particular, theoretical predictions regarding the non-Gaussian behavior of self-propelled particles are verified in the experiments. Furthermore, the full displacement probability distribution is analyzed, where in agreement with Brownian dynamics simulations either an extremely broadened peak or a pronounced double-peak structure is found depending on the experimental conditions.
\end{abstract}


\maketitle

\section{Introduction}
\label{sec:intro}
Recently, the single and collective properties of self-propelled particles have been studied intensely \cite{schimanskyreview,catesreview,marchettireview}.
Examples are found in quite different areas of physics and involve bacteria \cite{Swinney,2007SoEtAl,2004DoEtAl,2011Japan,Wensink_PNAS,Mino_Clement,Poon}, spermatozoa
\cite{2005Riedel_Science,DunkelPNAS,Woolley,Friedrich,2010Gompper}, and even fish, birds, and mammals including humans \cite{Vicsek_Report2012,Helbing,Schadschneider,Silverberg:13}.
Furthermore, various types of micron-sized man-made active particles have been developed \cite{Bibette,2011Herminghaus,Reinmueller,Sano_PRL2010,BechingerJPCM,snezhko_nature,2007Kapral,Sagues}.
One of the by now most popular artificial realizations of colloidal microswimmers are mesoscopic Janus particles which are put into motion by a chemical reaction in the surrounding solvent \cite{Paxton,ErbeBaraban}. 
In detail, this reaction is catalyzed at one surface of the Janus particle
such that an asymmetric gradient field arises, which self-propels the particle by diffusiophoresis \cite{Anderson,Golestanian_2007,Ismagilov,Dietrich}. Several features of the resulting swimming behavior such as the direction of motion \cite{Ke,Ebbens2011}, the dependence of the propulsion velocity on the particle size \cite{Golestanian_2012}, and the swimming efficiency \cite{SabassJCP} have been investigated recently. Focus has also been directed at the flow pattern in the vicinity of a heated Janus particle \cite{Bickel_PRE2013}, clustering in suspensions of self-propelled colloids \cite{Bialke_PRL2013,theurkauff2012dynamic,ChaikinScience,Baskaran_PRL2013}, and controlling the locomotion of single Janus micromotors \cite{Baraban_SoftMatter} by an external magnetic field \cite{Baraban_ACSnano,Baraban_2013}. Experiments with self-propelled spherical Janus particles in periodical arrangements of obstacles \cite{VolpeSM1} have inspired theoretical studies on possible applications for the sorting of chiral active particles \cite{VolpeSorting} or separation purposes in binary mixtures of passive colloids \cite{YangSeparation}. Very recent simulations of microswimmers moving in a ratchet channel also suggest their applicability for pumping processes \cite{Ghoshratchet}.
 
In general, the orientation of a Janus particle is fluctuating and therefore
the particle performs a persistent random walk \cite{Howse_2007}.
The mean square displacement hence crosses over from a ballistic regime, where the particle on average is self-propelled along its orientation, to a long-time diffusive behavior. The transition between these two regimes basically occurs at a timescale corresponding to the inverse rotational diffusion constant, i.e., when the particle has lost the memory of its initial orientation.
However, while the mean square displacement is the standard quantity to characterize modes
of propagation in self-propelled systems, there are only few studies for the  non-Gaussian behavior as revealed in the higher moments and in particular in the excess kurtosis.
Pure theoretical calculations \cite{Cond_Matt,tenHagen_JPCM} have addressed higher moments,
but an analysis  has never been performed based on experimental data for microswimmers. 
Non-Gaussianity is an important feature also in other disciplines of statistical physics including, e.g., the glass transition \cite{KobPRL,PuertasCates,BinderZippelius,Arbe02}
and the analysis of rare events (like earthquakes and stock crashes) \cite{bouchaud_book}.
Therefore, it is relevant from a fundamental point of view to get insight into the non-Gaussian statistics for microswimmer motion.

Here, we analyze higher moments characterizing non-Gaussianity in experimental trajectories of self-motile Janus particles and compare them to the theoretical predictions of a model based on the Langevin equations for the coupled translational and rotational motion of active Brownian particles. Moreover, we elucidate the interplay between the random and deterministic components of the particle displacements at very short times. 
We show that the crossover from diffusive short-time motion to super-diffusive motion at intermediate times \cite{tenHagen_JPCM} can also be verified experimentally, which supports the theoretical description of microswimmers by active Brownian models.
Additional insights regarding the non-Gaussianity are obtained by analyzing the time evolution of the full probability distribution of particle displacements. Here, the experimental data show that the initial Gaussian curve transforms into a shape with a significantly broadened peak if Janus particles in solutions with low hydrogen peroxide (H$_2$O$_2$) concentrations are considered. In contrast, for high H$_2$O$_2$ concentration a pronounced double-peak structure is found. These fundamentally different features result from a restriction of the rotational Brownian motion in the case of strongly driven Janus particles. Our observations are confirmed by Brownian dynamics simulations, where the particle orientation is either freely diffusing on a unit sphere or restrained to a two-dimensional plane.

This paper is organized as follows: section \ref{sec:methods} introduces the methods used in experiment, theory, and simulation. The experimental observations are presented in Sec.\ \ref{experiment}, where also a detailed discussion and interpretation in the context of the theoretical model is given. Finally, we conclude in Sec.\ \ref{sec:conc}.

\section{Methods}
\label{sec:methods}

\subsection{Experiment}
In our experiments, we study the motion of spherical Pt-silica Janus particles. The fabrication method is similar to that illustrated in Ref.\ \cite{Golestanian_2007}. By electron beam evaporation, a layer of Pt (thickness about $\unit{7}{\nano\metre}$) is deposited on the surface of one hemisphere of the particles (see Sec.\ \ref{sec:preparation} in the appendix for further details).
After that the Janus particles are resuspended in distilled water ($\unit{18.2}{\mega\ohm \centi\metre}$).
Most of the experiments are performed with Janus spheres with diameter $d_1=\unit{2.08 \pm 0.05}{\micro\metre}$ (measured by scanning electron microscopy).
Whenever additional results for smaller particles with diameter $d_2=\unit{0.96 \pm 0.03}{\micro\metre}$ are included for comparison, this is appropriately indicated.

The particle trajectories in water and in H$_2$O$_2$ solutions with different concentrations (1.25--$\unit{15}{\%}$) are observed by video microscopy with an image field of view of $512\times512$ pixels (approximately $\unit{80\times80}{\micro\metre}$). To be able to observe also the particle dynamics at very short times, the time interval $\Delta t$ between two images was reduced to $\unit{10}{\milli\second}$.
After the preparation of the solutions, a $\unit{70}{\micro\litre}$ droplet with specified H$_2$O$_2$ concentration was put on a cover slip. Image series consisting of 600--1000 frames were captured in one position located about  2--\unit{5}{\micro\metre} above the glass substrate.

In the same droplet, five movies were taken in five different locations in the same horizontal plane. The measurements for each H$_2$O$_2$ concentration were repeated in 12--15 droplets independently.
In order to have a good measurement reproducibility and to limit the influence of temperature and concentration
fluctuations induced by the chemical reaction, a fresh test solution for each droplet was reprepared. The experiments were performed in the stationary regime between $\unit{1}{\minute}$ and $\unit{9}{\minute}$ after the beginning of the catalytic reaction
in the H$_2$O$_2$ solution. In this period the fuel concentration does not change significantly since the used particle density is very low. The displacements of the Janus particles were measured by trajectory tracking from the movies. To reach the requirements of the statistical analysis, for each concentration more than 1000 particles were considered.

In the images the Janus particles appear half bright (the silica side) and half dark (Pt coating side). In order to determine the exact center of each Janus particle,
a two-step method using ``find edge'' and ``Gaussian blur'' was performed (see Sec.\ \ref{sec:processing} in the appendix for details). Thus, the center of the Janus particles could be determined with a $\pm0.5$ pixel accuracy. After this preprocessing, the trajectories of individual particles  can be tracked from the video material.

The dynamics of the particles in our system is strongly influenced by their translational and rotational Brownian motion. Thus, before investigating the self-propulsion on top of it, the diffusion coefficients $D_\mathrm{t}$ for translation and $D_\mathrm{r}$ for rotation have to be addressed. The translational diffusion coefficient is in principle given by the Stokes-Einstein equation
\begin{equation}
\label{Stokes}
D_\mathrm{t} = \frac{k_\mathrm{B} T}{3 \pi \eta d}\,,
\end{equation}
where $k_\mathrm{B} T$ is the thermal energy and $\eta$ is the viscosity of the solvent. Alternatively, $D_\mathrm{t}$ can also be directly determined from the two-dimensional mean square displacement $\left\langle (\Delta \mathbf r)^2\right\rangle$ of passive Brownian particles via $D_\mathrm{t} = \left\langle (\Delta \mathbf r)^2\right\rangle/(4 \Delta t)$. Following this standard method the experimental data yield $D_\mathrm{t}=\unit{0.175}{\micro\squaren\metre\reciprocal\second}$ for the particles with diameter $d_1=\unit{2.08}{\micro\metre}$ (theoretical prediction based on Eq.\ (\ref{Stokes}): $D_\mathrm{t}=\unit{0.211}{\micro\squaren\metre\reciprocal\second}$). In the case of the smaller particles $d_2=\unit{0.96}{\micro\metre}$ the measurements give $D_\mathrm{t}=\unit{0.416}{\micro\squaren\metre\reciprocal\second}$ as compared to the theoretical value $D_\mathrm{t}=\unit{0.456}{\micro\squaren\metre\reciprocal\second}$.
The small deviations between the measured and the predicted values are clearly due to hydrodynamic interactions with the glass substrate \cite{HappelB1991,Jeffrey1992}, which slightly reduce the mobility of the particles. We estimate the rotational diffusion coefficient from the relation $D_\mathrm{r}=3 D_\mathrm{t}/d^2$, which directly follows from Eq.\ (\ref{Stokes}) and its analogon
\begin{equation}
\label{rotdiffusion}
D_\mathrm{r} = \frac{k_\mathrm{B} T}{\pi \eta d^3}
\end{equation}
for rotational diffusion \cite{Dhont_book}.
Using the experimentally determined values for $D_\mathrm{t}$, one obtains $D_\mathrm{r}=\unit{0.121}{\reciprocal\second}$ for the larger particles and $D_\mathrm{r}=\unit{1.35}{\reciprocal\second}$ for the smaller ones.

\subsection{Theory}
\label{sec:theory}
In order to describe the dynamics of the Janus particles in our experiments, we use a theoretical model similar to that studied in detail in Ref.\ \cite{tenHagen_JPCM}. This general model for self-propelled Brownian particles is altered in a way such that it suits our experimental setup. Primarily, this means that the theoretical description is transferred from a one-particle situation to a dilute system with many, but not interacting particles as realized in our experiments. As the particles have different initial orientations that cannot easily be measured with sufficient accuracy, we always take an average and use corresponding theoretical results.

Starting with the Langevin equations for the overdamped motion of a Brownian particle, we include an effective driving force $\mathbf{F}= F \hat{\mathbf{u}}$, which accounts for the detailed self-propulsion mechanism of the active Janus particle on average and does not contradict the fact that the motion of a swimmer is force-free. $\mathbf{F}$ is parallel to a particle-fixed orientation vector $\hat{\mathbf{u}}$ that is defined by the position of the Pt layer (see Fig.\ \ref{Figanglesketch}).

The translational motion of the Janus spheres studied here is performed in two dimensions as gravity, in combination with electrostatic repulsion, keeps the particles close to the substrate, where the focal plane of the microscope is located.
However, in principle the particles can rotate freely. This implies that the translational motion of one Janus particle is described by the two-dimensional projection of the Langevin equation
\begin{equation}
\label{Langevinx1}
\frac{\mathrm{d}\mathbf{r}}{\mathrm{d}t} = \beta D_\mathrm{t} F \hat{\mathbf{u}} + \sqrt{2 D_\mathrm{t}} \boldsymbol{\xi}_{\mathbf{r}}
\end{equation}
for the center-of-mass position $\mathbf{r}(t)=(x(t),y(t))$, where $\beta = 1/(k_\mathrm{B} T)$ is the inverse effective thermal energy. As the direction of the self-propulsion depends on the particle orientation $\hat{\mathbf{u}}$, Eq.\ (\ref{Langevinx1}) is coupled to the rotational Langevin equation
\begin{equation}
\label{rotLangevin}
\frac{\mathrm{d}\hat{\mathbf{u}}}{\mathrm{d}t}=\sqrt{2 D_\mathrm{r}} \boldsymbol{\xi}_{\hat{\mathbf{u}}} \times \hat{\mathbf{u}}\,.
\end{equation}
The translational and rotational random motion due to the kicks of the solvent molecules is included by the Gaussian noise terms $\boldsymbol{\xi}_{\mathbf{r}}$ and $\boldsymbol{\xi}_{\hat{\mathbf{u}}}$ with zero mean and variances
$\langle\boldsymbol{\xi}_{\mathbf{r}}(t_{1})\otimes\boldsymbol{\xi}_{\mathbf{r}}(t_{2})\rangle = \langle\boldsymbol{\xi}_{\hat{\mathbf{u}}}(t_{1})\otimes\boldsymbol{\xi}_{\hat{\mathbf{u}}}(t_{2})\rangle
=\delta(t_{1}-t_{2}) \mathbbm{1}$, where $\mathbbm{1}$ is the unit tensor.
The corresponding orientational probability distribution for the freely diffusing orientation vector \cite{Dhont_book} is given by
\begin{equation}
\label{Entwicklungp2}
P(\theta, \varphi, t)  = \sum_{l=0}^\infty \sum_{m=-l}^l \mathrm{e}^{-D_\mathrm{r} l(l+1)t}\,Y_l^{m *}(\theta_0, \varphi_0) Y_l^m(\theta, \varphi)\,,
\end{equation}
where $Y_l^m$ and $Y_l^{m *}$ are the spherical harmonics and their complex conjugates. The spherical coordinates $\theta$ and  $\varphi$ define the particle orientation $\hat{\mathbf{u}}=(\sin \theta \sin \varphi, \sin \theta \cos \varphi, \cos \theta)$. Initial values at $t=0$ are indicated by the index 0.
For freely diffusing Janus particles with arbitrary initial orientation the analytical expressions for the different moments of the displacement probability distribution are given in Secs.\ \ref{sec:msd} and \ref{sec:kurtosis}. These results are in good agreement with the experimental data for up to \unit{5}{\%} H$_2$O$_2$ concentration.
As discussed in detail in Sec.\ \ref{sec:2d}, our observations strongly suggest that for higher H$_2$O$_2$ concentration of the solvent the particle orientation is not homogeneously distributed on a unit sphere, but is to some extent restricted to the two-dimensional plane of translational motion. This requires an appropriate adaption of the theoretical model.

\subsection{Simulation}
\label{sec:simulation}
\begin{figure}[tbh]
\centering
\includegraphics[width = \columnwidth]{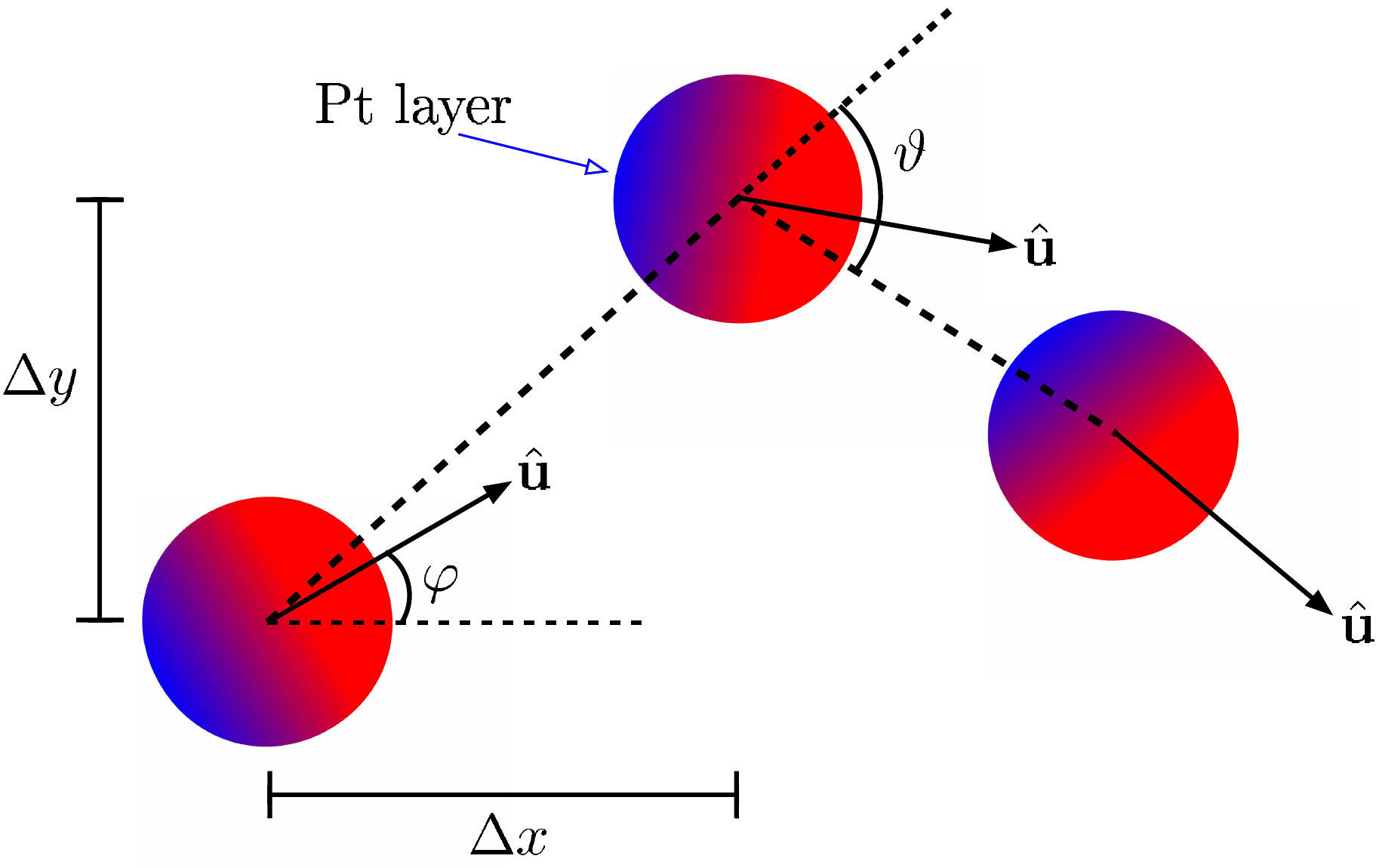}
\caption{\label{Figanglesketch}(Color online) Schematic of the particle motion for two subsequent time steps and definition of several parameters used for its characterization. The translational motion is determined by the displacements $\Delta x$ and $\Delta y$ of the center-of-mass postion of the particle. The orientation vector $\hat{\mathbf{u}}=(\sin \theta \sin \varphi, \sin \theta \cos \varphi, \cos \theta)$ coincides with the direction of self-propulsion. Note that $\theta = \unit{90}{\degree}$ in the figure for the sake of clarity. While $\theta$ and $\varphi$ define the particle orientation, $\vartheta$ is the angle between the directions of subsequent displacements. Due to the combination of Brownian motion and self-propulsion, $\hat{\mathbf{u}}$ is not necessarily collinear with the displacement direction.}
\end{figure}

While our model provides analytical expressions for the displacement moments, a corresponding Brownian dynamics simulation based on the same Langevin equations (\ref{Langevinx1}) and (\ref{rotLangevin}) allows us to study also the full distribution.
Numerical results are obtained for $10^{6}$ particle trajectories with arbitrary inital conditions and length 100 $t_\mathrm{r}$, where $t_\mathrm{r} = 1/D_\mathrm{r}$ is the rotational diffusion time.
The translational and rotational noise terms $\boldsymbol{\xi}_{\mathbf{r}}$ and $\boldsymbol{\xi}_{\hat{\mathbf{u}}}$ are implemented by independent Gaussian random numbers with zero mean and unit variance for each component.
Simulation results are provided for the probability distributions for both the magnitude and the direction of displacements. The function $\Psi(\Delta x,t)$
gives the probability to find a particle at a certain distance $\Delta x$ from its initial position after a specified time $t$ (see schematic illustration in Fig.\ \ref{Figanglesketch}). The time evolution of $\Psi(\Delta x,t)$ is discussed in detail in Sec.\ \ref{sec:pdf}.
To elucidate the interplay between the random and the deterministic components of the particle motion, we also address the probability distribution $\Psi(\vartheta,t)$ of the angle $\vartheta$ between the
directions of subsequent particle displacements (cf., Fig.\ \ref{Figanglesketch}) both in experiment and simulation (see Sec.\ \ref{sec:angle}).

\section{Results}
\label{experiment}

\subsection{Mean square displacement (MSD)}
\label{sec:msd}

\begin{figure}[tbhp]
\centering
\includegraphics[width = \columnwidth]{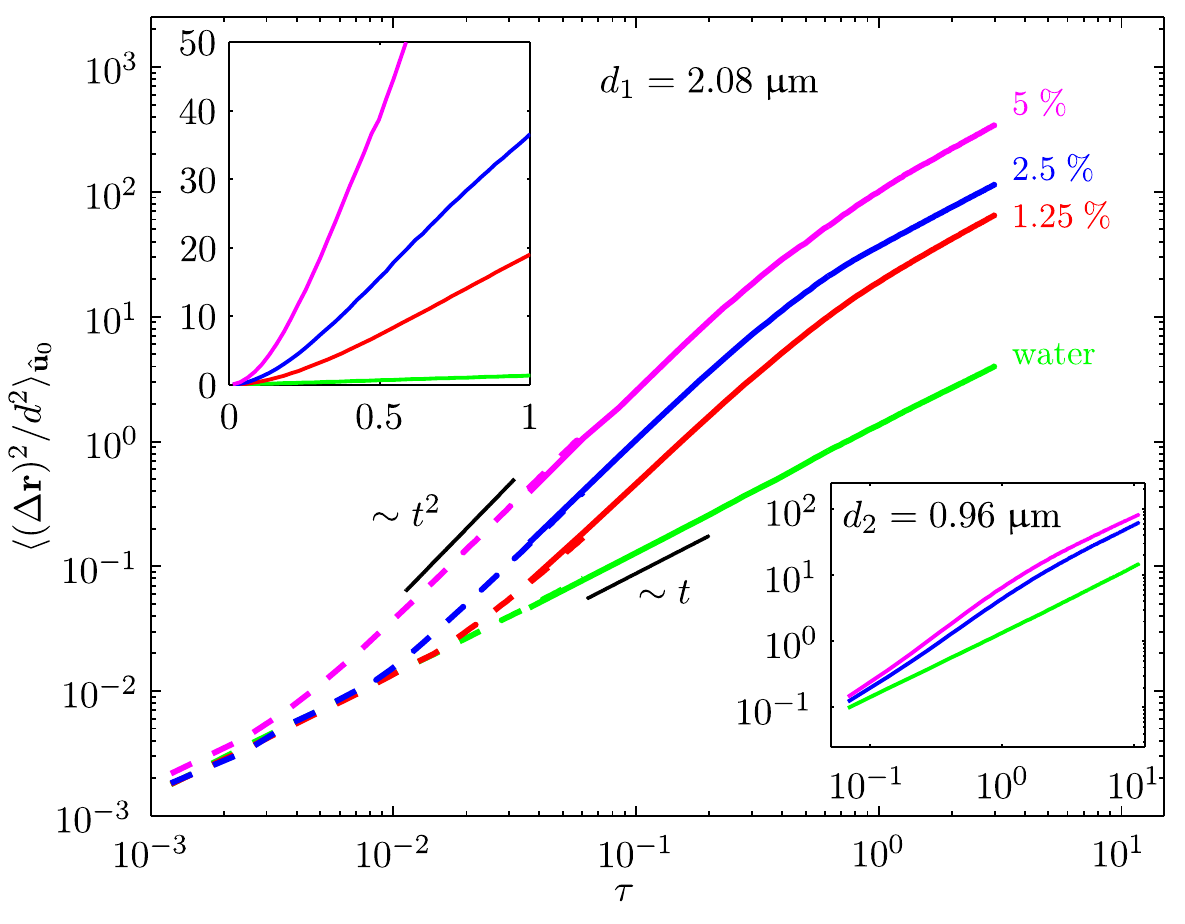}
\caption{\label{fig:expMSD}(Color online) Double logarithmic plot of the experimental results for the MSD of Janus particles with diameter $d_1=\unit{2.08}{\micro\metre}$ in water and in H$_2$O$_2$ solutions with different concentrations as function of the scaled time $\tau = D_\mathrm{r} t$.
Various regimes of motion are identified. Dashed and solid curves refer to different measurements for the same H$_2$O$_2$ concentrations. Left inset: visualization of the data in a linear plot. Right inset: experimental results for smaller particles with diameter $d_2=\unit{0.96}{\micro\metre}$ in water and in  H$_2$O$_2$ solutions with concentrations of \unit{2.5}{\%} and \unit{5}{\%}.}
\end{figure}

To characterize the dynamics of the Janus particles, we first discuss the MSD $\langle (\Delta \mathbf r)^2  \rangle_{\hat{\mathbf{u}}_0}$. Here, $\Delta \mathbf{r}= \mathbf{r}(t)-\mathbf{r}_0$ is the two-dimensional translational displacement and the notation $\langle ... \rangle_{\hat{\mathbf{u}}_0}$ denotes a noise average with an additional averaging over the initial orientation $\hat{\mathbf{u}}_0$ of the particles.
Figure \ref{fig:expMSD} shows the experimental results for the MSD in a double logarithmic plot. We use dimensionless quantities $\langle (\Delta \mathbf r)^2 \rangle_{\hat{\mathbf{u}}_0}/d^{2}$ and $\tau = D_\mathrm{r} t$ as this is convenient for the discussion of the measurements in the context of our theoretical model. While the main figure and the left inset of Fig.\ \ref{fig:expMSD} are based on measurements for particles with diameter $d_1 = \unit{2.08}{\micro\meter}$, the right inset visualizes corresponding data for smaller particles ($d_2=\unit{0.96}{\micro\meter}$). Due to the different rotational diffusion coefficients ($D_\mathrm{r}=\unit{0.121}{\reciprocal\second}$ for $d_1$ and $D_\mathrm{r}=\unit{1.35}{\reciprocal\second}$ for $d_2$),
the larger particles are more appropriate to study also the behavior at small values of $\tau$. Therefore, we focus on these particles for our detailed statistical analysis.
In the experiments, images were usually recorded with a frame rate of 10 frames per second (fps). Corresponding results for the MSD in water and in H$_2$O$_2$ solutions with different concentrations ranging from \unit{1.25}{\%} to \unit{5}{\%} are visualized by the solid lines in Fig.\ \ref{fig:expMSD}. However, to be able to resolve the very early time regime, additional measurements with a frame rate of \unit{100}{fps} are included as well (see dashed lines in Fig.\ \ref{fig:expMSD}).
In water the Janus particles undergo simple Brownian motion resulting in a linear time dependence of the MSD (see lowermost curve in Fig.\ \ref{fig:expMSD}). This changes when the particles are embedded in H$_2$O$_2$ solutions. A chemical reaction catalyzed by the Pt coated Janus particles is induced in the solvent \cite{Howse_2007}, which triggers the self-propulsion and leads to three different regimes of motion.

At short times ($\tau < 10^{-2}$ for \unit{2.5}{\%} H$_2$O$_2$ concentration), the particles undergo simple Brownian motion. The behavior corresponds to that of passive Brownian particles as the deterministic displacements due to the self-propulsion are not relevant at this early stage. We introduce the characteristic timescale $\tau_{1}$ to describe the transition to the intermediate regime, where directed (active) motion dominates.
Physically, $\tau_{1}$ is the time that is required for the chemical reaction to bring about a propulsive motion comparable to the Brownian random displacements. It clearly decreases with increasing H$_2$O$_2$ concentration of the solution and can be used to measure the strength of the self-propulsion of the investigated particles.

In the second regime, the MSD yields a super-diffusive behavior (approximately $\langle (\Delta \mathbf r)^2 \rangle_{\hat{\mathbf{u}}_0} \propto t^{2}$) as the motion is dominated by the directed propulsive component.
Finally, at a second timescale $\tau_2$ the dynamics becomes diffusive again with an enhanced diffusion coefficient \cite{Howse_2007}.
The transition to this third regime is also obvious in the linear plot of the MSD (see left inset in Fig.\ \ref{fig:expMSD}), where the nonlinear (quadratic) dependence at short times becomes linear for longer times.
The transition occurs near $\tau_2 = 1/2$, which corresponds to $t_2=\unit{4.1}{\second}$. This is the timescale where the particles lose their memory of the initial orientation due to rotational Brownian motion. Note that $\tau_2$ is largely independent of the H$_2$O$_2$ concentration as opposed to the transition time $\tau_1$.
Previous experiments \cite{Howse_2007,Ke} have observed the timescale $\tau_{2}$. However, an experimental investigation of the timescale $\tau_{1}$ has not been reported yet.

\begin{figure}[tbhp]
\centering
\includegraphics[width = \columnwidth]{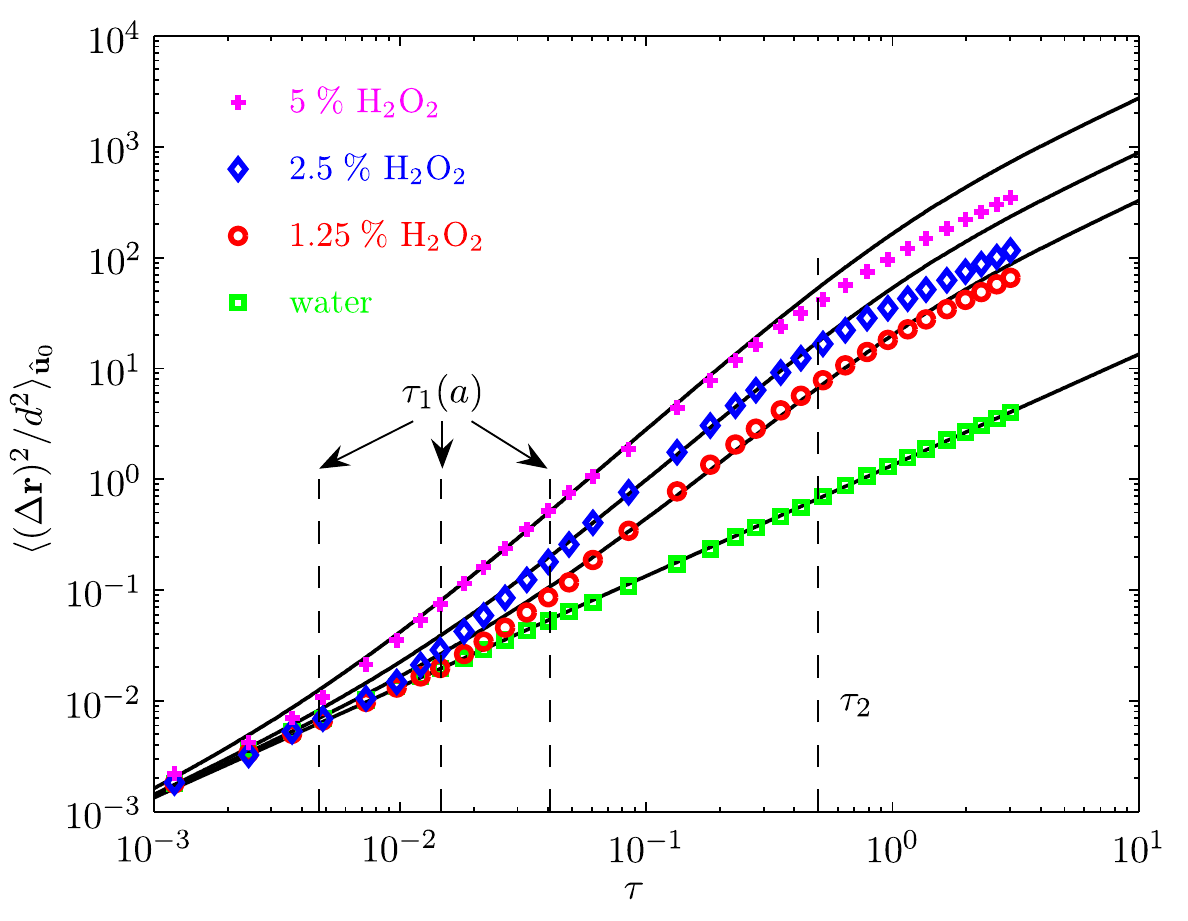}
\caption{\label{fig:theoMSD}(Color online) Comparison of the measured MSD (symbols) with the theoretical prediction (solid curves). The fitting parameter $a$ is given in Tab.\ \ref{tab:a}. Dashed lines indicate the transition times $\tau_1(a) = 18/a^2$ and $\tau_2 = 1/2$ between the different regimes of motion.}
\end{figure}

As visualized in Fig.\ \ref{fig:theoMSD}, the experimental data show good agreement with our theoretical model.
The solid curves represent best fits for short and intermediate times based on the prediction
\begin{equation}
\label{Moment2}
\left\langle \frac{(\Delta \mathbf r)^2}{d^2}\right\rangle_{\hat{\mathbf{u}}_0} = \frac{4}{3} \tau + \frac{1}{27} a^2   \biggl[2 \tau -1  +\mathrm{e}^{-2 \tau}  \biggr]
\end{equation}
for the two-dimensional MSD, which is obtained from Eqs.\ (\ref{Langevinx1}) and (\ref{Entwicklungp2}).
Here and in the following the dimensionless parameter $a = \beta d F$ is used to characterize the strength of the self-propulsion.
The fit curves in Fig.\ \ref{fig:theoMSD} are based on the values of $a$ specified in Tab.\ \ref{tab:a}.
We attribute the slight deviations at long times to small particle imperfections, in particular with regard to the Pt layer.
These might induce a non-central effective driving force, which leads to a tiny, but deterministic rotation of the particle and thus reduces the measured MSD for long times.
This effect could be included in the theoretical model either by means of a renormalized rotational diffusion coefficient \cite{Howse_2007}
 or by explicitly considering an internal torque generated by the asymmetry of the particle \cite{Kuemmel:13}.
Another source of deviations might be remnants of long-ranged hydrodynamic effects. Short-ranged particle-particle interactions can be excluded due to our tracking algorithm, where only particles with a specified minimum distance from each other are considered (for further details see Sec.\ \ref{sec:processing} in the appendix).
From Eq.\ (\ref{Moment2}) one also obtains a prediction for the transition time $\tau_1$ between the initial diffusive and the super-diffusive regime. Equating the Brownian and the propulsive contributions yields $\tau_1 = 18/a^2$. In agreement with the experimental observations, $\tau_1$ is antiproportional to the square of the self-propulsion force.

\begin{table}[ht]
\caption{\label{tab:a}Dimensionless self-propulsion force $a = \beta d F$ of the Janus particles as a function of the H$_2$O$_2$ concentration of the solvent.  The values for $a$ are obtained by fitting Eq.\ (\ref{Moment2}) to the experimental data for the MSD (see Fig.\ \ref{fig:theoMSD}).}
\begin{ruledtabular}
\begin{tabular}{cc}
H$_2$O$_2$ concentration [\%] & scaled self-propulsion $a$ \\
\hline
0 & 0\\
  1.25 & 21\\
  2.5 & 35\\
  5 & 62\\
 \end{tabular}
\end{ruledtabular}
\end{table}

\subsection{Excess kurtosis}
\label{sec:kurtosis}
\begin{figure}[tbh]
\centering
\includegraphics[width = \columnwidth]{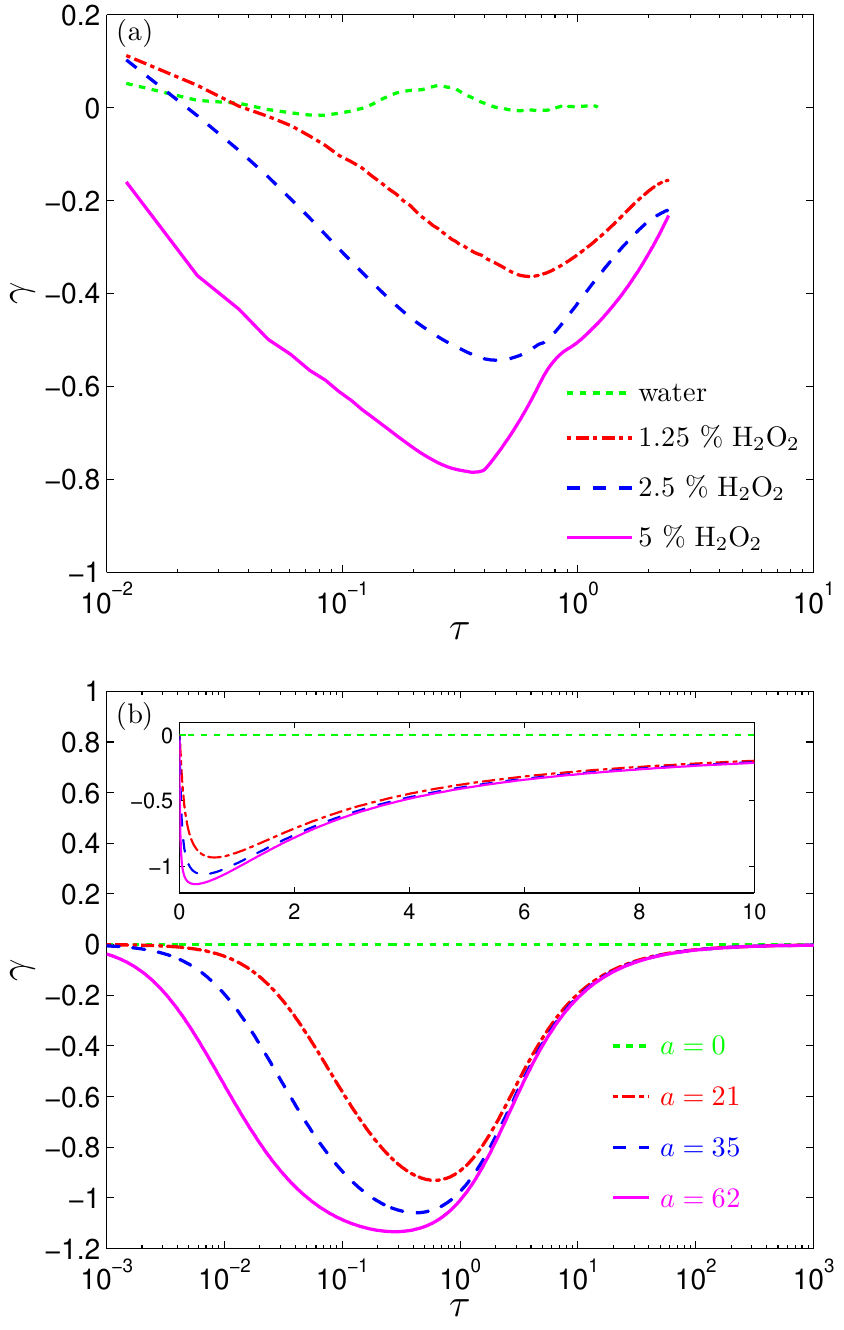}
\caption{\label{FigKurt}(Color online) (a) Experimental and (b) theoretical results for the excess kurtosis $\gamma$. The theoretical curves are calculated for the values of the self-propulsion force extracted from the MSD fits in Fig.\ \ref{fig:theoMSD} (see Tab.\ \ref{tab:a}). Inset in (b): Visualization of the theoretical data in a linear plot}
\end{figure}

On top of the analysis of the MSD, here we also address skewness $S$ and excess kurtosis $\gamma$, which serve to quantify the non-Gaussian behavior of self-propelled particles \cite{Cond_Matt}. They are given by
\begin{equation}
\label{skewness}
S=\frac{\left\langle(\Delta x)^3\right\rangle_{\hat{\mathbf{u}}_0}}{\left\langle(\Delta x)^2\right\rangle_{\hat{\mathbf{u}}_0}^{3/2}}
\end{equation}
and
\begin{equation}
\label{kurtosis}
\gamma=\frac{\left\langle(\Delta x)^4\right\rangle_{\hat{\mathbf{u}}_0}}{\left\langle(\Delta x)^2\right\rangle_{\hat{\mathbf{u}}_0}^2}-3\,,
\end{equation}
respectively. Note that Eqs.\ (\ref{skewness}) and (\ref{kurtosis}) are only valid because $\left\langle\Delta x\right\rangle_{\hat{\mathbf{u}}_0}=0$ in our system. Otherwise, the moments have to be replaced by the respective central moments. As the third moment $\langle(\Delta x)^3\rangle_{\hat{\mathbf{u}}_0}$ trivially vanishes due to the symmetry of $\Psi(\Delta x,t)$, resulting from the averaging over the initial orientation $\hat{\mathbf{u}}_0$ of the Janus particles, the skewness $S$ is zero. Though, our measurements in H$_2$O$_2$ solution clearly yield nonzero values for the excess kurtosis $\gamma$, which directly indicates non-Gaussian behavior. The curves in Fig.\ \ref{FigKurt}(a) are calculated from the experimental displacement data based on Eq.\ (\ref{kurtosis}).
Results are shown for pure water and H$_2$O$_2$ concentrations of \unit{1.25}{\%}, \unit{2.5}{\%}, and \unit{5}{\%} corresponding to the analysis of the MSD in Figs.\ \ref{fig:expMSD} and \ref{fig:theoMSD}. As expected, the reference measurements in water yield a nearly vanishing excess kurtosis $\gamma$, which indicates largely Gaussian behavior.
The slight deviations from zero can be induced by a not perfectly symmetric particle shape. This leads to a situation similar to the Brownian motion of passive ellipsoids, where also small positive values for the non-Gaussian parameter are observed \cite{Han:06}.
However, the time dependence of the excess kurtosis changes drastically, when active Janus particles in H$_2$O$_2$ solutions are considered.
The measured curves turn negative and present a minimum located between $\tau = 0.4$ and $\tau = 0.8$ depending on the H$_2$O$_2$ concentration. If the latter is increased, the position of the minimum is shifted to shorter times and it becomes more pronounced ($\gamma_\mathrm{min} \approx -0.35 $ for \unit{1.25}{\%} and $\gamma_\mathrm{min} \approx -0.8$ for \unit{5}{\%} H$_2$O$_2$). This corresponds to the general observation that for all times a higher H$_2$O$_2$ concentration leads to more negative values for $\gamma$.

To derive the analytical expression for the excess kurtosis from our theoretical model, in addition to the MSD also the fourth moment $\langle(\Delta x)^4\rangle_{\hat{\mathbf{u}}_0}$ of $\Psi(\Delta x,t)$ is required. For the situation in our experiments, where the particles undergo three-dimensional rotational Brownian motion, one obtains
	\begin{equation}
	\label{Moment4}
	\begin{split}
\left\langle \frac{(\Delta x)^4}{d^4}\right\rangle_{\hat{\mathbf{u}}_0} & =
 {\frac {4}{3}}\,{\tau}^{2}    +{\frac {2}{27}}\,a^{2}\tau \left[2 \tau -1 + \mathrm{e}^{-2 \tau} \right]  \\
&\quad  +{\frac {1}{21870}}\,a^{4}\Bigl[90 \tau^{2} - 156 \tau + 107  \\
&\quad  -54 \tau \mathrm{e}^{-2 \tau} - 108 \mathrm{e}^{-2 \tau} +\mathrm{e}^{-6 \tau} \Bigr]\,.
\end{split}
\end{equation}
The final result for the excess kurtosis $\gamma$ directly follows from Eq.\ (\ref{kurtosis}) by inserting Eq.\ (\ref{Moment4}) and $\left\langle(\Delta x)^2\right\rangle_{\hat{\mathbf{u}}_0} = (1/2)\left\langle(\Delta \mathbf{r})^2\right\rangle_{\hat{\mathbf{u}}_0}$ (see Eq.\ (\ref{Moment2})).
In Fig.\ \ref{FigKurt}(b) theoretical curves are plotted for $a$ as determined for pure water and the various H$_2$O$_2$ concentrations from the analysis of the MSD (see Fig.\ \ref{fig:theoMSD} and Tab.\ \ref{tab:a}). The linear plot in the inset visualizes the pronounced negative long-time tail \cite{tenHagen_JPCM}.

Basically, the theoretical results show the same tendency as discussed for the experimental curves in Fig.\ \ref{FigKurt}(a).
In particular with regard to the general behavior and the position of the minimum the agreement is very good, although the experimental values for $\gamma$ are usually less negative than the theoretical predictions. Slightly positive values as measured for very short times can again be ascribed to small deviations from an ideal isotropic particle shape, similar to our observations in pure water.

\begin{figure}[tbh]
\centering
\includegraphics[width =\columnwidth]{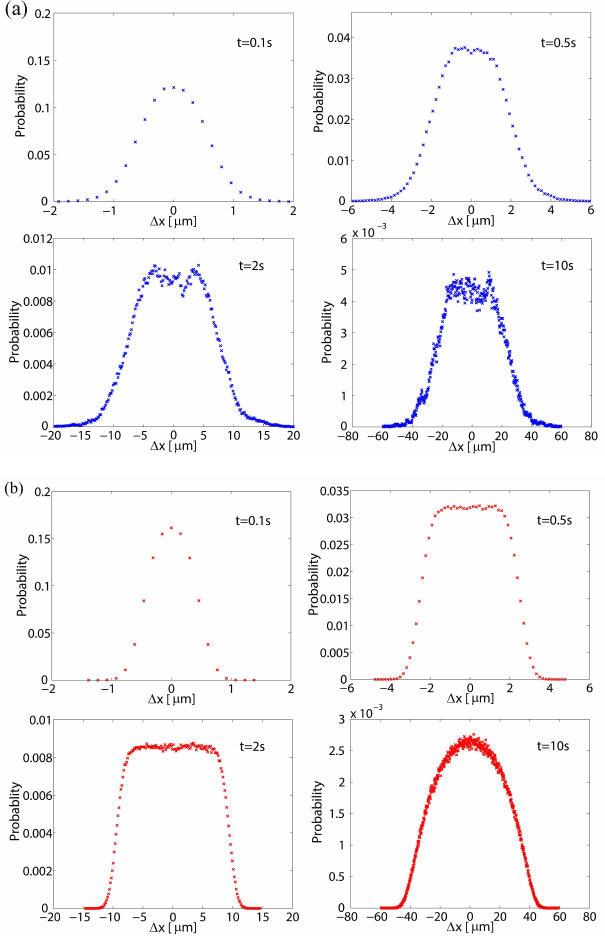}
\caption{\label{Fig322}(Color online) Time evolution of $\Psi(\Delta x,t)$: (a) experimental results for $\unit{5}{\%}$ H$_2$O$_2$ concentration, (b) corresponding simulation for $a=62$.}
\end{figure}

\subsection{Displacement probability distribution}
\label{sec:pdf}

After the discussion of the displacement moments, in a next step we study the full probability distribution function
$\Psi(\Delta x,t)$ for one-dimensional displacements, which reveals further details of the statistical characteristics of the particle motion.
In water, the Janus particles show a simple diffusive behavior corresponding to Gaussian probability distributions at all times.
However, for self-propelled particles the curves for $\Psi(\Delta x,t)$ significantly deviate from a Gaussian shape. In Fig.\ \ref{Fig322}(a) exemplarily the experimental results for \unit{5}{\%} H$_2$O$_2$ concentration are given. Data points are plotted for each pixel, corresponding to an interval of $\unit{0.16}{\micro\metre}$.
At the beginning ($t=\unit{0.1}{\second}$),
$\Psi(\Delta x,t)$  is still nearly Gaussian. After $t=\unit{0.5}{\second}$, a broadening of the peak is observed, which further intensifies until $t=\unit{2}{\second}$. Furthermore, the wings of the distribution become steeper as time proceeds.

A theoretical prediction for $\Psi(\Delta x,t)$ is  obtained numerically from the model equations (\ref{Langevinx1}) and (\ref{rotLangevin}).
As opposed to the analytical results for the MSD and the excess kurtosis presented in Secs.\ \ref{sec:msd} and \ref{sec:kurtosis}, the full displacement probability distribution is only accessibe via a Brownian dynamics simulation (see Sec.\ \ref{sec:simulation}).
Figure \ref{Fig322}(b) gives the simulation results calculated for $a = \beta d F = 62$. They show the same characteristic features -- such as the broadened peak and the steep wings -- as the experimental plots.

The shape of the displacement probability distribution curves is closely related to the particle dynamics in the different regimes of motion (see Sec.\ \ref{sec:msd}). At short times, when the random translational motion still dominates, $\Psi(\Delta x,t)$  is nearly Gaussian. In the intermediate regime, where the self-propulsion dictates the particle motion, the broadening of the peak emerges (see Fig.\ \ref{Fig322}, plots for  $t=\unit{0.5}{\second}$ ($\tau = 0.0605$) and $t=\unit{2}{\second}$ ($\tau = 0.242$)). Thus, the appearance of the broadened peak accompanied by the steep wings is due to the active component of the motion. This shape also provides the explanation for the negative values of the excess kurtosis $\gamma$ (see Sec.\ \ref{sec:kurtosis}), which could only be suspected in earlier theoretical calculations \cite{Cond_Matt,tenHagen_JPCM}.

As we assume that the Janus particles undergo free rotational Brownian motion in three dimensions, their initial orientations are homogeneously distributed on a unit sphere \cite{Marsaglia}. This implies that the projections of all possible initial orientation vectors $\hat{\mathbf{u}}_0$ on the $x$ axis are evenly spread between minus one and one. Consequently, the contribution to the deterministic particle displacement in $x$ direction is uniformly distributed as well, which explains the kind of rectangular shape of $\Psi(\Delta x,t)$ in the intermediate regime.
Although the measured MSD (see Fig.\ \ref{fig:expMSD}) already indicates diffusive behavior again for $\tau > 1/2$, the non-Gaussian structure of $\Psi(\Delta x,t)$ still persists (see Fig.\ \ref{Fig322}, plots for $t=\unit{10}{\second}$ ($\tau = 1.21$)). This yields that the displacement probability distribution is less sensitive to changes in the type of motion than the MSD. The prolonged presence of the broadened peak is consistent with the negative long-time tail observed for the excess kurtosis (see Fig.\ \ref{FigKurt}) and explains its origin.
For very long times, $\Psi(\Delta x,t)$ is expected to become Gaussian again. While the experiments cannot be performed long enough to show this tendency clearly, it is confirmed by our simulation. The conversion back to a Gaussian shape occurs at $\tau$ on the order of $10^2$, when also the excess kurtosis, which is a direct measure for the non-Gaussianity, approaches zero again.

\subsection{Directional probability distribution}
\label{sec:angle}
In this section, we briefly discuss an alternative approach to visualize the relative importance of the random and the deterministic contributions to the particle motion. It is based on the probability distribution function $\Psi(\vartheta,t)$ for the angle $\vartheta$ between the directions of subsequent particle displacements (see Fig.\ \ref{Figanglesketch}). While the Brownian noise induces arbitrary displacement directions (corresponding to a homogeneous distribution of $\vartheta$ between $-\pi$ and $\pi$), the self-propulsive motion is always collinear with the particle orientation $\hat{\mathbf{u}}$, determined by the position of the Pt layer, and thus favors values of $\vartheta$ near zero.

\begin{figure*}[tbp]
\centering
\includegraphics[width = 1.6\columnwidth]{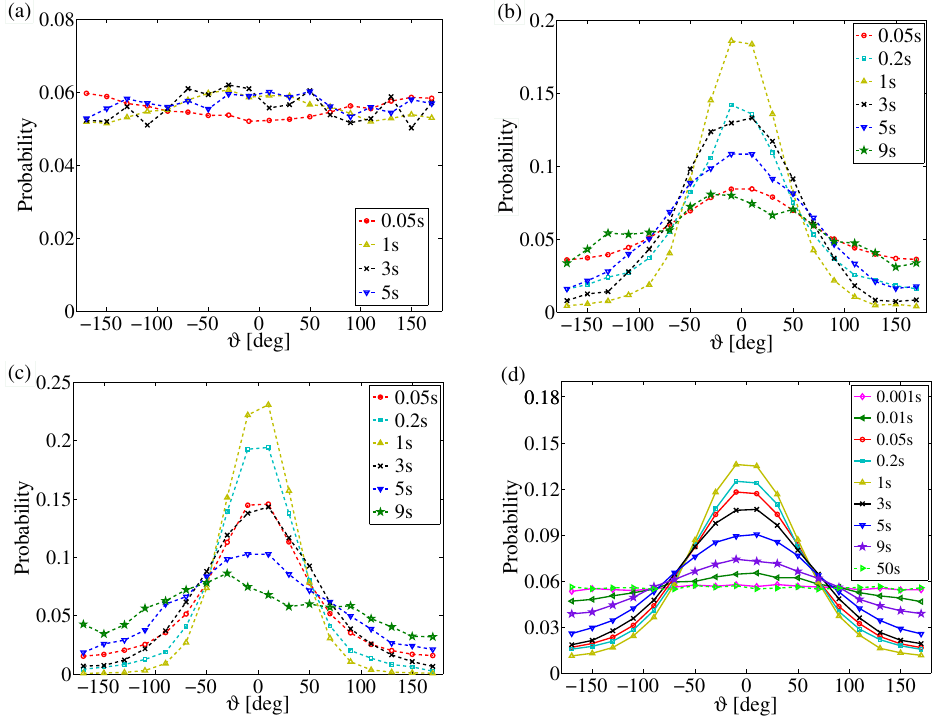}
\caption{\label{Fig34}(Color online) Time evolution of $\Psi(\vartheta,t)$: (a)--(c) experimental data for (a) water, (b) \unit{2.5}{\%},  and (c) \unit{5}{\%} H$_2$O$_2$ concentration. (d) Simulation results for $a=62$.}
\end{figure*}

The experimental results for the time evolution of $\Psi(\vartheta,t)$ in solutions with different H$_2$O$_2$ concentrations are shown in Fig.\ \ref{Fig34}(a)--(c).
In water (see Fig.\ \ref{Fig34}(a)) the distribution is uniform at all times due to the random Brownian motion. However, for nonzero H$_2$O$_2$ concentration a peaked behavior of $\Psi(\vartheta,t)$ occurs (see Figs.\ \ref{Fig34}(b) for \unit{2.5}{\%} and \ref{Fig34}(c) for \unit{5}{\%} H$_2$O$_2$). Here, the peak height increases for short times until it reaches its maximum value at about \unit{1}{\second}. After that the curves become flatter again when the displacement directions decorrelate due to rotational Brownian motion.
With increasing H$_2$O$_2$ concentration the peak attains higher maximum values and it becomes more pronounced at intermediate times. At long times there is no significant difference between the curves for \unit{2.5}{\%} and \unit{5}{\%} H$_2$O$_2$ concentration.

Figure \ref{Fig34}(d) gives the simulation results for the time evolution of $\Psi(\vartheta,t)$ for $a=62$. It is in good agreement with the corresponding experimental data and shows additional curves for very short and very long times that are not directly accessible in experiment. The three regimes (short-time diffusive, intermediate ballistic, and long-time diffusive) discussed in detail in the previous sections can also be extracted from the plots of $\Psi(\vartheta,t)$. For very short times (see curves for $t=\unit{0.001}{\second}$ and $t=\unit{0.01}{\second}$ in Fig.\ \ref{Fig34}(d)) the directions of the particle displacements in two adjacent time intervals are completely uncorrelated. This yields that the passive Brownian motion is dominant in this regime. The pronounced peaks for intermediate values of $t$ (see in particular curves for $t=\unit{0.5}{\second}$ and $t=\unit{1}{\second}$ in Fig.\ \ref{Fig34}(d)) clearly show that the particle dynamics is largely influenced by the directed self-propelling component of the motion. Finally, for very long times the angular probability distribution becomes homogeneous again, indicating the long-time diffusive regime.

\subsection{Orientational symmetry breaking for high H$_2$O$_2$ concentration}
\label{sec:2d}
\begin{figure}[tbh]
\centering
\includegraphics[width = \columnwidth]{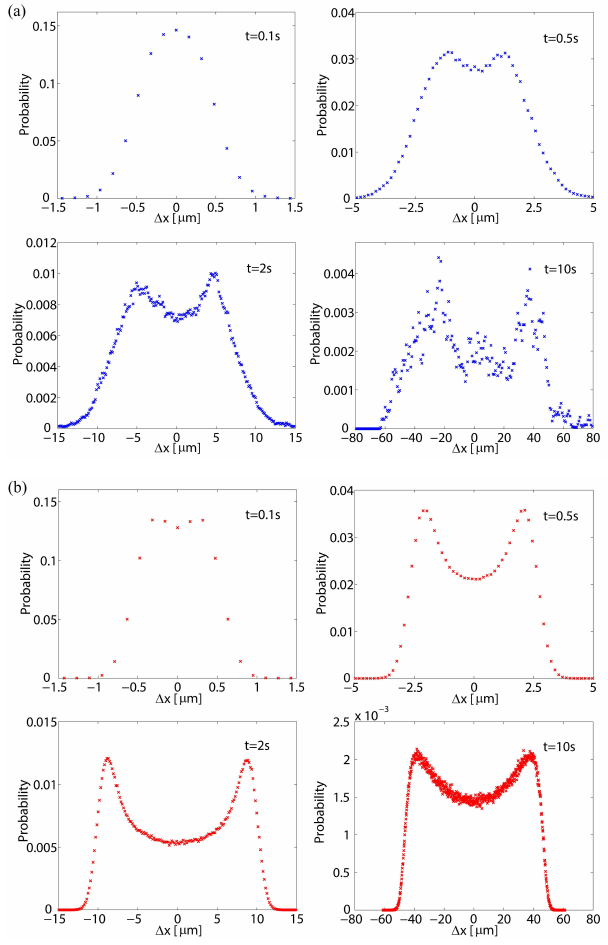}
\caption{\label{doublepeak}(Color online) (a) Time evolution of the measured probability distribution $\Psi(\Delta x,t)$ for Janus particles in a $\unit{10}{\%}$ H$_2$O$_2$ solution. The occurrence of the double peak indicates that the particle orientation does not diffuse freely on a unit sphere for high H$_2$O$_2$ concentrations. (b) Reference simulation for particles whose orientation is restricted to the $x$-$y$ plane.}
\end{figure}
The previous discussion focused on results for up to \unit{5}{\%} H$_2$O$_2$ concentration. We have also performed experiments with \unit{10}{\%} and \unit{15}{\%} solutions. Here, our video observation of the Janus particles strongly indicates that their orientation is not freely diffusing on a unit sphere any more, but is largely restricted to the $x$-$y$ plane.
This symmetry breaking in the rotational motion directly affects the structure of the probability distribution function $\Psi(\Delta x,t)$ and also leads to different analytical expressions for the displacement moments. Assuming that the orientation vector of the particle lies always inside the two-dimensional plane of motion, the evolution of the single orientational angle $\phi$ is given by \cite{tenHagen_JPCM}
\begin{equation}
\label{Gaussianphi}
P(\phi ,t)=\frac{1}{\sqrt{4\pi D_\mathrm{r}t}}\exp\left({-\frac{(\phi-\phi_0)^2}{4D_\mathrm{r}t}}\right)\,.
\end{equation}
Consequently, from Eqs.\ (\ref{Langevinx1}) and (\ref{Gaussianphi}) one obtains the orientation-averaged MSD
\begin{equation}
\label{2dMoment2}
\left\langle \frac{(\Delta \mathbf r)^2}{d^2}\right\rangle_{\hat{\mathbf{u}}_0} = \frac{4}{3} \tau + \frac{2}{9} a^2   \biggl[\tau -1  +\mathrm{e}^{-\tau}  \biggr]
\end{equation}
and the fourth moment
\begin{equation}
\label{2dMoment4}
\begin{split}
\left\langle \frac{(\Delta x)^4}{d^4}\right\rangle_{\hat{\mathbf{u}}_0} & =
 {\frac {4}{3}}\,{\tau}^{2}    +{\frac {4}{9}}\,a^{2}\tau \left[\tau -1 + \mathrm{e}^{-\tau} \right]  \\
&\quad  +{\frac {1}{3888}}\,a^{4}\Bigl[144 \tau^{2} - 540 \tau + 783  \\
&\quad  -240 \tau \mathrm{e}^{- \tau} - 784 \mathrm{e}^{-\tau} +\mathrm{e}^{-4 \tau} \Bigr]
\end{split}
\end{equation}
determining the excess kurtosis.
At first sight, these results seem to be very similar to their counterparts for free three-dimensional rotational Brownian motion as presented in Eqs.\ (\ref{Moment2}) and (\ref{Moment4}), respectively. Technically, they only differ in the  prefactors of the various terms and in the arguments of the exponential functions. The larger absolute values of the latter for three-dimensional orientation indicate that the particles lose their orientational memory earlier than in the case with two-dimensional rotational Brownian motion. Despite the formal analogy of the analytical expressions for the displacement moments, both the experimental data and the simulation results reveal striking differences with regard to the full probability distribution function (see Fig.\ \ref{doublepeak}). While an extremely broadened peak is observed for isotropic rotational diffusion (see Fig.\ \ref{Fig322}), a characteristic double peak occurs due to the symmetry breaking that restricts the particle orientations to the two-dimensional plane of translational motion (see Fig.\ \ref{doublepeak}). It is most pronounced after times on the order of several seconds.
The origin of the double peak can be understood by considering the initial orientations of the Janus particles. If these are homogeneously distributed on a unit circle (and not on a unit sphere), the corresponding projections on the $x$ axis are not evenly spread between minus one and one. Instead of that, values close to the extrema have a higher statistical weight than values around zero. Consequently, the majority of the Janus particles carry out a significant directed displacement during the super-diffusive regime where the self-propulsion is dominant. Only few particles stay close to their initial position. This explains the characteristic double-peak structure observed in our experiments and verified by a corresponding computer simulation (see Fig.\ \ref{doublepeak}).

\begin{figure}[tbp]
\centering
\includegraphics[width = \columnwidth]{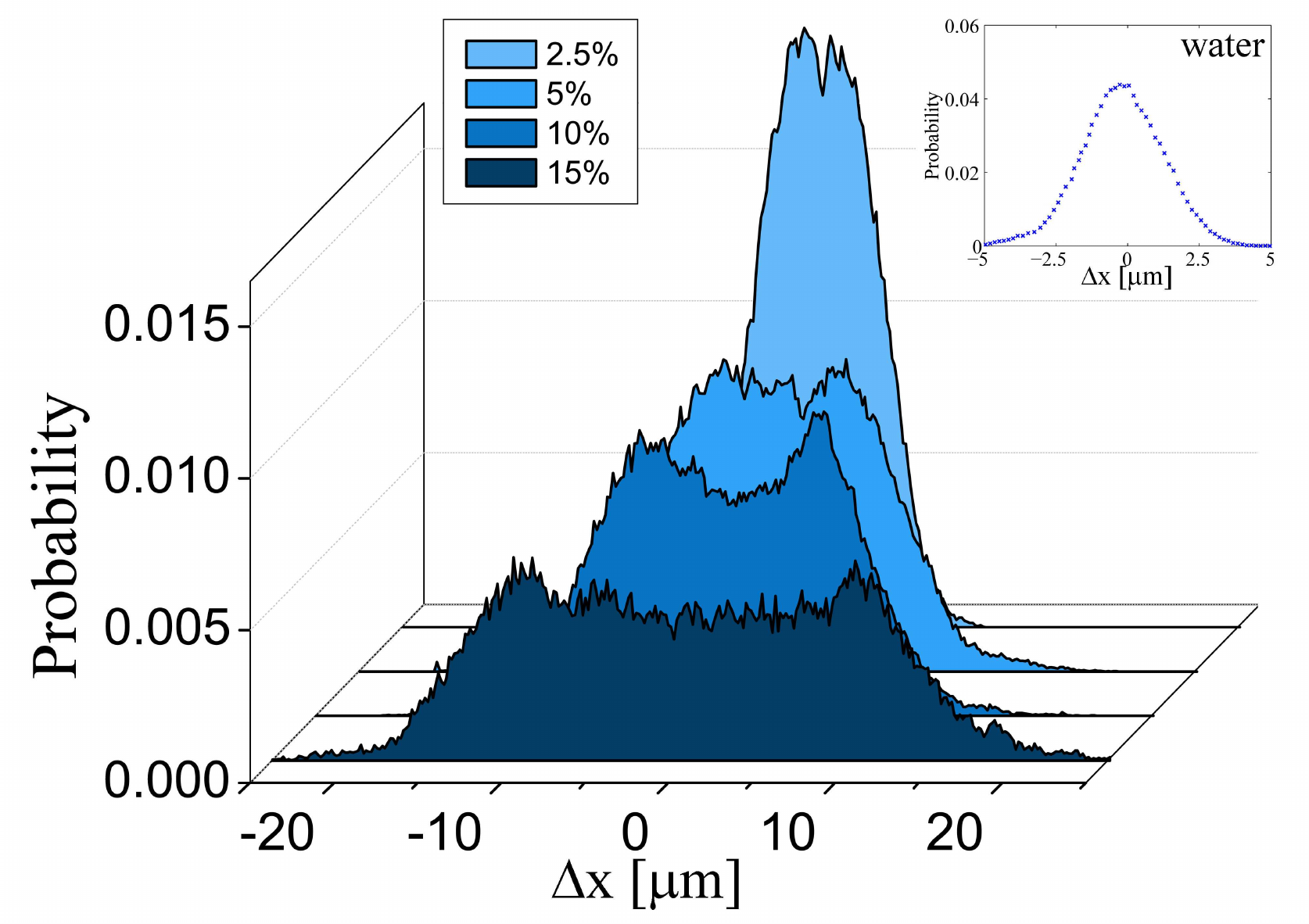}
\caption{\label{Fig321}(Color online) Experimental results for $\Psi(\Delta x,t)$ after \unit{2}{\second} for different H$_2$O$_2$ concentrations.
The inset shows the Gaussian distribution measured in water.}
\end{figure}

Figure \ref{Fig321} directly visualizes the dependence of $\Psi(\Delta x,t)$  on the H$_2$O$_2$ concentration. For this purpose, snapshots of the distributions after $\unit{2}{\second}$ are shown. These reveal Gaussian behavior for pure water, a broadened peak for low, and a double peak for high H$_2$O$_2$ concentration.
The existence of the double peak in the latter case is a second independent indicator for the orientational symmetry breaking, in addition to our video observation.
We surmise that the limitation of the rotational freedom is due to hydrodynamic effects \cite{Gauger:06,Goetze:10}. In solutions with higher H$_2$O$_2$ concentration, the chemical reaction generates a stronger  self-propulsion. Thus, the flow pattern in the vicinity of the Janus particles \cite{Bickel_PRE2013} might have increasing influence on the rotational motion. However, clearly more work is needed to fully understand the origin of the observed orientational symmetry breaking.

Finally, our theoretical description including limited rotational freedom could also explain the seemingly contradicting experimental results presented in Refs.\ \cite{Ke} and \cite{Howse_2007}.  In Ref.\ \cite{Howse_2007} the rotational diffusion time $t_\mathrm{r} = 1/D_\mathrm{r}$ is measured to decrease as a function of the H$_2$O$_2$ concentration, which is attributed to an asymmetric Pt coverage leading to a deterministic rotation of the particles.  On the contrary, a slight increase of $t_\mathrm{r}$ with higher H$_2$O$_2$ concentration is reported in Ref.\ \cite{Ke}, where $t_\mathrm{r}$ is estimated from the transition between the super-diffusive and the long-time diffusive regime.
Following the argument of limited rotational freedom, this increase is not due to a change of the rotational diffusion coefficient, but could directly be explained by the different prefactors of $\tau$ in the exponents of Eqs.\ (\ref{Moment2}) and (\ref{2dMoment2}).  The real situation in experiments with active Janus particles is most likely always somewhere in between free rotational diffusion and full restriction to two dimensions. While the good agreement between theory and experiment for low H$_2$O$_2$ concentrations (up to \unit{5}{\%}) -- as discussed in Secs.\ \ref{sec:msd}-\ref{sec:angle} -- implies that the orientational limitation plays a minor role in those cases, a modified description is required for higher H$_2$O$_2$ concentrations.

\section{Conclusions}
\label{sec:conc}

In summary, we have studied the non-Gaussian characteristics of the diffusiophoretic motion of self-propelled Pt-silica Janus spheres both in experiment and in theory. The propulsion strength is varied by means of different concentrations of the H$_2$O$_2$ solution, in which the particles are embedded.  
The good agreement between theory and experiment shows that in spite of the rather complicated underlying propulsion mechanism all the main features of the motion including the higher displacement moments can be understood by our model based on the translational and orientational Langevin equations \cite{Dhont_book,tenHagen_JPCM}. The analytical predictions have been experimentally verified not only for the mean square displacement, but also for the excess kurtosis characterizing the non-Gaussian behavior. This promises the applicability of our model to a broad range of experimental systems as the detailed propulsion mechanism can be accounted for by the implementation of an effective driving force.
As illustrated here, the excess kurtosis is a helpful tool beyond the standard mean square displacement approach in order to understand the interplay between the deterministic and the random components of the dynamics of active Brownian systems. The characteristic non-Gaussian super-diffusive intermediate regime is enframed by two diffusive regimes -- simple (passive) Brownian motion at short times and enhanced diffusion with a significantly increased diffusion constant \cite{Howse_2007} due to the active part of the motion at long times.

A deeper understanding of the non-Gaussianity is provided by the full probability distributions for the magnitude and the direction of displacements as obtained from the experiments in good agreement with a corresponding Brownian dynamics simulation.
Concerning the magnitude of the displacements, the respective probability distribution for low H$_2$O$_2$ concentration reveals a significantly broadened peak at intermediate times, which is induced by the self-propulsion of the Janus particles. In agreement with the negative long-time tail of the excess kurtosis, the broadened peak is still observable when the particle dynamics has already changed to the enhanced diffusive regime.
This phenomenon can be traced back to the super-diffusive regime, where a large number of particles performed significant deterministic displacements. In the experiments with high H$_2$O$_2$ concentration, a symmetry breaking manifested in a limitation of the rotational Brownian motion is found. It induces a pronounced double-peak structure of the displacement probability distribution and requires a modification of the theoretical description.

In order to generalize the presented results for spherical Janus particles, in a next step, it is interesting to analyze the non-Gaussian behavior of asymmetric particles. These can either be axisymmetric such as rods \cite{HoeflingPRL} and ellipsoids \cite{Han:09}, or they can have an even more complicated anisotropic shape \cite{WittkowskiL2012,Kraft:13}. While some results for the non-Gaussian behavior of passive \cite{Han:06} and active \cite{tenHagen_JPCM} axisymmetric particles  are already available, an open question addresses the influence of more complicated particle shapes on the characteristic features of the particle dynamics beyond simple Brownian motion. In particular, an additional torque \cite{vanTeeffelenL2008} -- as automatically induced by an asymmetry around the propulsion axis \cite{Kuemmel:13} -- significantly affects the motional behavior and leads to a modified displacement probability distribtion. Another interesting aspect for future experimental studies are solvent flow effects \cite{Holzer:10,Zoettl:12} which accelerate the displacement of microswimmers drastically \cite{tenHagenPRE:11}.
In the present work the non-Gaussianity is already caused by the presence of the self-propulsion of the active particles. Thus, here it is a single particle phenomenon as dilute systems, where particle interaction are negligible, are investigated in our experiments. However, for situations with higher particle density \cite{Wensink2008,WensinkJPCM,KaiserPRE}, the interplay between hydrodynamic effects \cite{Lauga_review:09}
and the self-propulsion of the particles is expected to give rise to new physical phenomena manifested also in the excess kurtosis of the displacement probability distribution and its higher moments.

\acknowledgments
We thank Jiang Lei for using electron beam evaporation to prepare the Janus particles.
This work was financially supported by the National Natural Science Foundation of China (No. 11272322, 11202219, and 21005058), the ERC Advanced Grant INTERCOCOS (Grant No. 267499), and by the DFG within SFB TR6 (project C3).

\appendix
\section{Experimental apparatus and methods}
\label{appendix}

\subsection{Preparation of the Janus particles}
\label{sec:preparation}
\begin{figure}[tbh]
\centering
\includegraphics[width = \columnwidth]{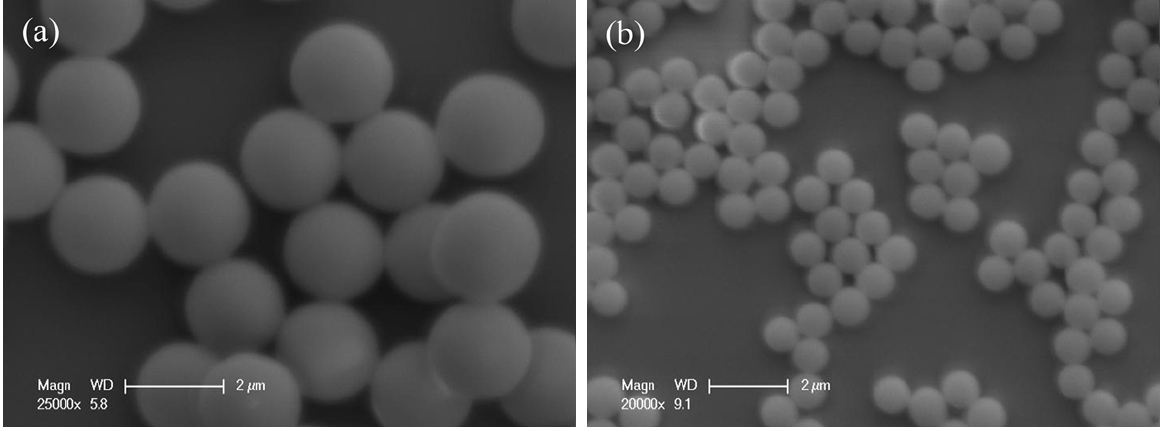}
\caption{\label{FigSEM}SEM images of the silica particles with diameters (a) $d_1 = \unit{2.08 \pm 0.05}{\micro\metre}$ and (b) $ d_2 =\unit{0.96 \pm 0.03}{\micro\metre}$.}
\end{figure}

The silica particles used in the experiments were produced
by the University of Petroleum in China.
The diameters of the two considered particle sizes are $d_1=\unit{2.08 \pm 0.05}{\micro\metre}$  and  $d_2=\unit{0.96 \pm 0.03}{\micro\metre}$
measured by scanning electron microscopy (SEM) (see Fig.\ \ref{FigSEM}).

To fabricate the Janus particles, an aqueous suspension of silica particles is first deposited on
a 4-inch silicon wafer by spin coating at low speed (\unit{800}{rpm}). After evaporating the water,
a single layer of particles is formed on the wafer. Then, using electron
beam evaporation (by an Innotec e-beam evaporator in the Institute of Semiconductors, Chinese Academy of Sciences),
a layer of Pt (thickness about $\unit{7}{\nano\metre}$) is deposited on the upper half surfaces of the particles.
Finally, the half-coated Janus particles are collected from the silicon wafer using a razor blade
and resuspended in distilled water ($\unit{18.2}{\mega\ohm \centi\metre}$). The volumetric concentration of the Janus particle suspension is approximately $5 \times 10^{-3}$.

\subsection{Image processing}
\label{sec:processing}
\begin{figure}[tbh]
\centering
\includegraphics[width = \columnwidth]{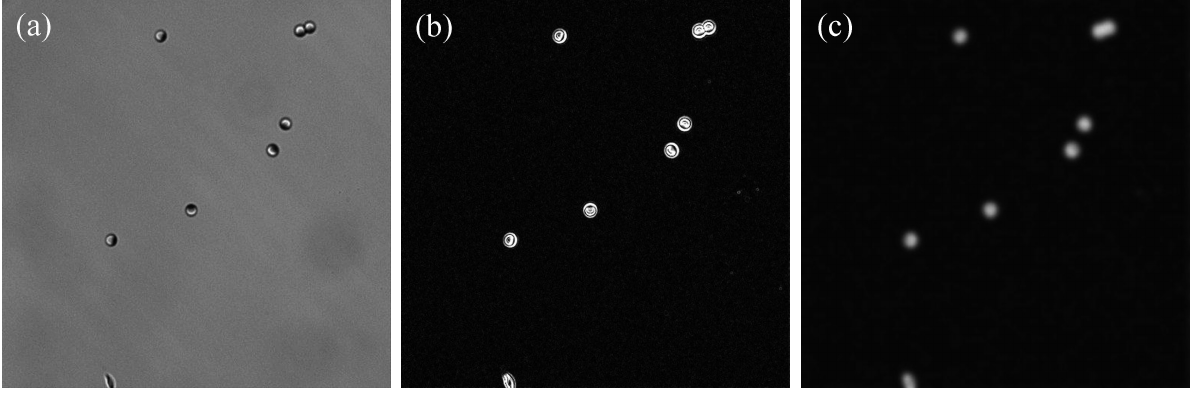}
\caption{\label{Figimaging}Image preprocessing with the program \textit{ImageJ}: (a) the original image directly obtained by video microscopy from the experiments,
(b) image after using the ``find edge''
function, and (c) image after using the ``Gaussian blur'' function.
}
\end{figure}

We apply the following method to determine the exact center of the Janus particles, which appear half bright and half dark in the images (see Fig.\ \ref{Figimaging}). First, the ``find edge''
function of the program \textit{ImageJ} is used, which highlights sharp intensity changes. As the sharpest
changes occur at the particle edges, this function offers a way to reconstruct
the round shape of the particle. Secondly, using the ``Gaussian blur'' function of \textit{ImageJ}, the grayscale value distribution
in the particle domain is determined. The point with the maximum grayscale value is considered to be
the center of the particle. This method has a $\pm 0.5$ pixel accuracy.

After this preprocessing, the particle positions $(x,y)$ can be tracked by the software Video Spot Tracker (V07.02).
To guarantee that only individual particles are tracked, we omit aggregated particles and use a ``dead zone''
function, by which the region approximately one diameter around the particle is monitored. If other particles enter into this
``dead zone'', the tracking of the respective particles is stopped. Therefore, particle aggregation as well as particle-particle collisions and interactions
can be excluded from our investigation.

\bibliography{refs}

\begin{thebibliography}{77}%
\makeatletter
\providecommand \@ifxundefined [1]{%
 \@ifx{#1\undefined}
}%
\providecommand \@ifnum [1]{%
 \ifnum #1\expandafter \@firstoftwo
 \else \expandafter \@secondoftwo
 \fi
}%
\providecommand \@ifx [1]{%
 \ifx #1\expandafter \@firstoftwo
 \else \expandafter \@secondoftwo
 \fi
}%
\providecommand \natexlab [1]{#1}%
\providecommand \enquote  [1]{``#1''}%
\providecommand \bibnamefont  [1]{#1}%
\providecommand \bibfnamefont [1]{#1}%
\providecommand \citenamefont [1]{#1}%
\providecommand \href@noop [0]{\@secondoftwo}%
\providecommand \href [0]{\begingroup \@sanitize@url \@href}%
\providecommand \@href[1]{\@@startlink{#1}\@@href}%
\providecommand \@@href[1]{\endgroup#1\@@endlink}%
\providecommand \@sanitize@url [0]{\catcode `\\12\catcode `\$12\catcode
  `\&12\catcode `\#12\catcode `\^12\catcode `\_12\catcode `\%12\relax}%
\providecommand \@@startlink[1]{}%
\providecommand \@@endlink[0]{}%
\providecommand \url  [0]{\begingroup\@sanitize@url \@url }%
\providecommand \@url [1]{\endgroup\@href {#1}{\urlprefix }}%
\providecommand \urlprefix  [0]{URL }%
\providecommand \Eprint [0]{\href }%
\@ifxundefined \urlstyle {%
  \providecommand \doi  [0]{\begingroup \@sanitize@url \@doi}%
  \providecommand \@doi [1]{\endgroup \@@startlink {\doibase
  #1}doi:\discretionary {}{}{}#1\@@endlink }%
}{%
  \providecommand \doi  [0]{doi:\discretionary{}{}{}\begingroup
  \urlstyle{rm}\Url }%
}%
\providecommand \doibase [0]{http://dx.doi.org/}%
\providecommand \Doi [0]{\begingroup \@sanitize@url \@Doi }%
\providecommand \@Doi  [1]{\endgroup\@@startlink{\doibase#1}\@@Doi}%
\providecommand \@@Doi [1]{#1\@@endlink}%
\providecommand \selectlanguage [0]{\@gobble}%
\providecommand \bibinfo  [0]{\@secondoftwo}%
\providecommand \bibfield  [0]{\@secondoftwo}%
\providecommand \translation [1]{[#1]}%
\providecommand \BibitemOpen [0]{}%
\providecommand \bibitemStop [0]{}%
\providecommand \bibitemNoStop [0]{.\EOS\space}%
\providecommand \EOS [0]{\spacefactor3000\relax}%
\providecommand \BibitemShut  [1]{\csname bibitem#1\endcsname}%
\bibitem [{\citenamefont {Romanczuk}\ \emph {et~al.}(2012)\citenamefont
  {Romanczuk}, \citenamefont {B{\"a}r}, \citenamefont {Ebeling}, \citenamefont
  {Lindner},\ and\ \citenamefont {Schimansky-Geier}}]{schimanskyreview}%
  \BibitemOpen
  \bibfield  {author} {\bibinfo {author} {\bibfnamefont {P.}~\bibnamefont
  {Romanczuk}}, \bibinfo {author} {\bibfnamefont {M.}~\bibnamefont {B{\"a}r}},
  \bibinfo {author} {\bibfnamefont {W.}~\bibnamefont {Ebeling}}, \bibinfo
  {author} {\bibfnamefont {B.}~\bibnamefont {Lindner}}, \ and\ \bibinfo
  {author} {\bibfnamefont {L.}~\bibnamefont {Schimansky-Geier}},\ }\href@noop
  {} {\bibfield  {journal} {\bibinfo  {journal} {Eur. Phys. J. Special
  Topics},\ }\textbf {\bibinfo {volume} {202}},\ \bibinfo {pages} {1} (\bibinfo
  {year} {2012})}\BibitemShut {NoStop}%
\bibitem [{\citenamefont {Cates}(2012)}]{catesreview}%
  \BibitemOpen
  \bibfield  {author} {\bibinfo {author} {\bibfnamefont {M.~E.}\ \bibnamefont
  {Cates}},\ }\href@noop {} {\bibfield  {journal} {\bibinfo  {journal} {Rep.
  Prog. Phys.},\ }\textbf {\bibinfo {volume} {75}},\ \bibinfo {pages} {042601}
  (\bibinfo {year} {2012})}\BibitemShut {NoStop}%
\bibitem [{\citenamefont {Marchetti}\ \emph {et~al.}(2013)\citenamefont
  {Marchetti}, \citenamefont {Joanny}, \citenamefont {Ramaswamy}, \citenamefont
  {Liverpool}, \citenamefont {Prost}, \citenamefont {Rao},\ and\ \citenamefont
  {Simha}}]{marchettireview}%
  \BibitemOpen
  \bibfield  {author} {\bibinfo {author} {\bibfnamefont {M.~C.}\ \bibnamefont
  {Marchetti}}, \bibinfo {author} {\bibfnamefont {J.~F.}\ \bibnamefont
  {Joanny}}, \bibinfo {author} {\bibfnamefont {S.}~\bibnamefont {Ramaswamy}},
  \bibinfo {author} {\bibfnamefont {T.~B.}\ \bibnamefont {Liverpool}}, \bibinfo
  {author} {\bibfnamefont {J.}~\bibnamefont {Prost}}, \bibinfo {author}
  {\bibfnamefont {M.}~\bibnamefont {Rao}}, \ and\ \bibinfo {author}
  {\bibfnamefont {R.~A.}\ \bibnamefont {Simha}},\ }\Doi
  {10.1103/RevModPhys.85.1143} {\bibfield  {journal} {\bibinfo  {journal} {Rev.
  Mod. Phys.},\ }\textbf {\bibinfo {volume} {85}},\ \bibinfo {pages} {1143}
  (\bibinfo {year} {2013})}\BibitemShut {NoStop}%
\bibitem [{\citenamefont {Chen}\ \emph {et~al.}(2012)\citenamefont {Chen},
  \citenamefont {Dong}, \citenamefont {Be'er}, \citenamefont {Swinney},\ and\
  \citenamefont {Zhang}}]{Swinney}%
  \BibitemOpen
  \bibfield  {author} {\bibinfo {author} {\bibfnamefont {X.}~\bibnamefont
  {Chen}}, \bibinfo {author} {\bibfnamefont {X.}~\bibnamefont {Dong}}, \bibinfo
  {author} {\bibfnamefont {A.}~\bibnamefont {Be'er}}, \bibinfo {author}
  {\bibfnamefont {H.~L.}\ \bibnamefont {Swinney}}, \ and\ \bibinfo {author}
  {\bibfnamefont {H.~P.}\ \bibnamefont {Zhang}},\ }\Doi
  {10.1103/PhysRevLett.108.148101} {\bibfield  {journal} {\bibinfo  {journal}
  {Phys. Rev. Lett.},\ }\textbf {\bibinfo {volume} {108}},\ \bibinfo {pages}
  {148101} (\bibinfo {year} {2012})}\BibitemShut {NoStop}%
\bibitem [{\citenamefont {Sokolov}\ \emph {et~al.}(2007)\citenamefont
  {Sokolov}, \citenamefont {Aranson}, \citenamefont {Kessler},\ and\
  \citenamefont {Goldstein}}]{2007SoEtAl}%
  \BibitemOpen
  \bibfield  {author} {\bibinfo {author} {\bibfnamefont {A.}~\bibnamefont
  {Sokolov}}, \bibinfo {author} {\bibfnamefont {I.~S.}\ \bibnamefont
  {Aranson}}, \bibinfo {author} {\bibfnamefont {J.~O.}\ \bibnamefont
  {Kessler}}, \ and\ \bibinfo {author} {\bibfnamefont {R.~E.}\ \bibnamefont
  {Goldstein}},\ }\Doi {10.1103/PhysRevLett.98.158102} {\bibfield  {journal}
  {\bibinfo  {journal} {Phys. Rev. Lett.},\ }\textbf {\bibinfo {volume} {98}},\
  \bibinfo {pages} {158102} (\bibinfo {year} {2007})}\BibitemShut {NoStop}%
\bibitem [{\citenamefont {Dombrowski}\ \emph {et~al.}(2004)\citenamefont
  {Dombrowski}, \citenamefont {Cisneros}, \citenamefont {Chatkaew},
  \citenamefont {Goldstein},\ and\ \citenamefont {Kessler}}]{2004DoEtAl}%
  \BibitemOpen
  \bibfield  {author} {\bibinfo {author} {\bibfnamefont {C.}~\bibnamefont
  {Dombrowski}}, \bibinfo {author} {\bibfnamefont {L.}~\bibnamefont
  {Cisneros}}, \bibinfo {author} {\bibfnamefont {S.}~\bibnamefont {Chatkaew}},
  \bibinfo {author} {\bibfnamefont {R.~E.}\ \bibnamefont {Goldstein}}, \ and\
  \bibinfo {author} {\bibfnamefont {J.~O.}\ \bibnamefont {Kessler}},\ }\Doi
  {10.1103/PhysRevLett.93.098103} {\bibfield  {journal} {\bibinfo  {journal}
  {Phys. Rev. Lett.},\ }\textbf {\bibinfo {volume} {93}},\ \bibinfo {pages}
  {098103} (\bibinfo {year} {2004})}\BibitemShut {NoStop}%
\bibitem [{\citenamefont {Ishikawa}\ \emph {et~al.}(2011)\citenamefont
  {Ishikawa}, \citenamefont {Yoshida}, \citenamefont {Ueno}, \citenamefont
  {Wiedeman}, \citenamefont {Imai},\ and\ \citenamefont
  {Yamaguchi}}]{2011Japan}%
  \BibitemOpen
  \bibfield  {author} {\bibinfo {author} {\bibfnamefont {T.}~\bibnamefont
  {Ishikawa}}, \bibinfo {author} {\bibfnamefont {N.}~\bibnamefont {Yoshida}},
  \bibinfo {author} {\bibfnamefont {H.}~\bibnamefont {Ueno}}, \bibinfo {author}
  {\bibfnamefont {M.}~\bibnamefont {Wiedeman}}, \bibinfo {author}
  {\bibfnamefont {Y.}~\bibnamefont {Imai}}, \ and\ \bibinfo {author}
  {\bibfnamefont {T.}~\bibnamefont {Yamaguchi}},\ }\Doi
  {10.1103/PhysRevLett.107.028102} {\bibfield  {journal} {\bibinfo  {journal}
  {Phys. Rev. Lett.},\ }\textbf {\bibinfo {volume} {107}},\ \bibinfo {pages}
  {028102} (\bibinfo {year} {2011})}\BibitemShut {NoStop}%
\bibitem [{\citenamefont {Wensink}\ \emph {et~al.}(2012)\citenamefont
  {Wensink}, \citenamefont {Dunkel}, \citenamefont {Heidenreich}, \citenamefont
  {Drescher}, \citenamefont {Goldstein}, \citenamefont {L\"owen},\ and\
  \citenamefont {Yeomans}}]{Wensink_PNAS}%
  \BibitemOpen
  \bibfield  {author} {\bibinfo {author} {\bibfnamefont {H.~H.}\ \bibnamefont
  {Wensink}}, \bibinfo {author} {\bibfnamefont {J.}~\bibnamefont {Dunkel}},
  \bibinfo {author} {\bibfnamefont {S.}~\bibnamefont {Heidenreich}}, \bibinfo
  {author} {\bibfnamefont {K.}~\bibnamefont {Drescher}}, \bibinfo {author}
  {\bibfnamefont {R.~E.}\ \bibnamefont {Goldstein}}, \bibinfo {author}
  {\bibfnamefont {H.}~\bibnamefont {L\"owen}}, \ and\ \bibinfo {author}
  {\bibfnamefont {J.~M.}\ \bibnamefont {Yeomans}},\ }\Doi
  {10.1073/pnas.1202032109} {\bibfield  {journal} {\bibinfo  {journal} {Proc.
  Natl. Acad. Sci. USA},\ }\textbf {\bibinfo {volume} {109}},\ \bibinfo {pages}
  {14308} (\bibinfo {year} {2012})}\BibitemShut {NoStop}%
\bibitem [{\citenamefont {Mi\~no}\ \emph {et~al.}(2011)\citenamefont {Mi\~no},
  \citenamefont {Mallouk}, \citenamefont {Darnige}, \citenamefont {Hoyos},
  \citenamefont {Dauchet}, \citenamefont {Dunstan}, \citenamefont {Soto},
  \citenamefont {Wang}, \citenamefont {Rousselet},\ and\ \citenamefont
  {Clement}}]{Mino_Clement}%
  \BibitemOpen
  \bibfield  {author} {\bibinfo {author} {\bibfnamefont {G.}~\bibnamefont
  {Mi\~no}}, \bibinfo {author} {\bibfnamefont {T.~E.}\ \bibnamefont {Mallouk}},
  \bibinfo {author} {\bibfnamefont {T.}~\bibnamefont {Darnige}}, \bibinfo
  {author} {\bibfnamefont {M.}~\bibnamefont {Hoyos}}, \bibinfo {author}
  {\bibfnamefont {J.}~\bibnamefont {Dauchet}}, \bibinfo {author} {\bibfnamefont
  {J.}~\bibnamefont {Dunstan}}, \bibinfo {author} {\bibfnamefont
  {R.}~\bibnamefont {Soto}}, \bibinfo {author} {\bibfnamefont {Y.}~\bibnamefont
  {Wang}}, \bibinfo {author} {\bibfnamefont {A.}~\bibnamefont {Rousselet}}, \
  and\ \bibinfo {author} {\bibfnamefont {E.}~\bibnamefont {Clement}},\ }\Doi
  {10.1103/PhysRevLett.106.048102} {\bibfield  {journal} {\bibinfo  {journal}
  {Phys. Rev. Lett.},\ }\textbf {\bibinfo {volume} {106}},\ \bibinfo {pages}
  {048102} (\bibinfo {year} {2011})}\BibitemShut {NoStop}%
\bibitem [{\citenamefont {Schwarz-Linek}\ \emph {et~al.}(2012)\citenamefont
  {Schwarz-Linek}, \citenamefont {Valeriani}, \citenamefont {Cacciuto},
  \citenamefont {Cates}, \citenamefont {Marenduzzo}, \citenamefont {Morozov},\
  and\ \citenamefont {Poon}}]{Poon}%
  \BibitemOpen
  \bibfield  {author} {\bibinfo {author} {\bibfnamefont {J.}~\bibnamefont
  {Schwarz-Linek}}, \bibinfo {author} {\bibfnamefont {C.}~\bibnamefont
  {Valeriani}}, \bibinfo {author} {\bibfnamefont {A.}~\bibnamefont {Cacciuto}},
  \bibinfo {author} {\bibfnamefont {M.~E.}\ \bibnamefont {Cates}}, \bibinfo
  {author} {\bibfnamefont {D.}~\bibnamefont {Marenduzzo}}, \bibinfo {author}
  {\bibfnamefont {A.~N.}\ \bibnamefont {Morozov}}, \ and\ \bibinfo {author}
  {\bibfnamefont {W.~C.~K.}\ \bibnamefont {Poon}},\ }\Doi
  {10.1073/pnas.1116334109} {\bibfield  {journal} {\bibinfo  {journal} {Proc.
  Natl. Acad. Sci. USA},\ }\textbf {\bibinfo {volume} {109}},\ \bibinfo {pages}
  {4052} (\bibinfo {year} {2012})}\BibitemShut {NoStop}%
\bibitem [{\citenamefont {Riedel}\ \emph {et~al.}(2005)\citenamefont {Riedel},
  \citenamefont {Kruse},\ and\ \citenamefont {Howard}}]{2005Riedel_Science}%
  \BibitemOpen
  \bibfield  {author} {\bibinfo {author} {\bibfnamefont {I.~H.}\ \bibnamefont
  {Riedel}}, \bibinfo {author} {\bibfnamefont {K.}~\bibnamefont {Kruse}}, \
  and\ \bibinfo {author} {\bibfnamefont {J.}~\bibnamefont {Howard}},\ }\Doi
  {10.1126/science.1110329} {\bibfield  {journal} {\bibinfo  {journal}
  {Science},\ }\textbf {\bibinfo {volume} {309}},\ \bibinfo {pages} {300}
  (\bibinfo {year} {2005})}\BibitemShut {NoStop}%
\bibitem [{\citenamefont {Kantsler}\ \emph {et~al.}(2013)\citenamefont
  {Kantsler}, \citenamefont {Dunkel}, \citenamefont {Polin},\ and\
  \citenamefont {Goldstein}}]{DunkelPNAS}%
  \BibitemOpen
  \bibfield  {author} {\bibinfo {author} {\bibfnamefont {V.}~\bibnamefont
  {Kantsler}}, \bibinfo {author} {\bibfnamefont {J.}~\bibnamefont {Dunkel}},
  \bibinfo {author} {\bibfnamefont {M.}~\bibnamefont {Polin}}, \ and\ \bibinfo
  {author} {\bibfnamefont {R.~E.}\ \bibnamefont {Goldstein}},\ }\href@noop {}
  {\bibfield  {journal} {\bibinfo  {journal} {Proc. Natl. Acad. Sci. USA},\
  }\textbf {\bibinfo {volume} {110}},\ \bibinfo {pages} {1187} (\bibinfo {year}
  {2013})}\BibitemShut {NoStop}%
\bibitem [{\citenamefont {Woolley}(2003)}]{Woolley}%
  \BibitemOpen
  \bibfield  {author} {\bibinfo {author} {\bibfnamefont {D.~M.}\ \bibnamefont
  {Woolley}},\ }\Doi {10.1530/rep.0.1260259} {\bibfield  {journal} {\bibinfo
  {journal} {Reproduction},\ }\textbf {\bibinfo {volume} {216}},\ \bibinfo
  {pages} {259} (\bibinfo {year} {2003})}\BibitemShut {NoStop}%
\bibitem [{\citenamefont {Friedrich}\ and\ \citenamefont
  {J\"ulicher}(2008)}]{Friedrich}%
  \BibitemOpen
  \bibfield  {author} {\bibinfo {author} {\bibfnamefont {B.~M.}\ \bibnamefont
  {Friedrich}}\ and\ \bibinfo {author} {\bibfnamefont {F.}~\bibnamefont
  {J\"ulicher}},\ }\Doi {10.1088/1367-2630/10/12/123025} {\bibfield  {journal}
  {\bibinfo  {journal} {New J. Phys.},\ }\textbf {\bibinfo {volume} {10}},\
  \bibinfo {pages} {123035} (\bibinfo {year} {2008})}\BibitemShut {NoStop}%
\bibitem [{\citenamefont {Elgeti}\ \emph {et~al.}(2010)\citenamefont {Elgeti},
  \citenamefont {Kaupp},\ and\ \citenamefont {Gompper}}]{2010Gompper}%
  \BibitemOpen
  \bibfield  {author} {\bibinfo {author} {\bibfnamefont {J.}~\bibnamefont
  {Elgeti}}, \bibinfo {author} {\bibfnamefont {U.~B.}\ \bibnamefont {Kaupp}}, \
  and\ \bibinfo {author} {\bibfnamefont {G.}~\bibnamefont {Gompper}},\ }\Doi
  {10.1016/j.bpj.2010.05.015} {\bibfield  {journal} {\bibinfo  {journal}
  {Biophys. J.},\ }\textbf {\bibinfo {volume} {99}},\ \bibinfo {pages} {1018}
  (\bibinfo {year} {2010})}\BibitemShut {NoStop}%
\bibitem [{\citenamefont {Vicsek}\ and\ \citenamefont
  {Zafeiris}(2012)}]{Vicsek_Report2012}%
  \BibitemOpen
  \bibfield  {author} {\bibinfo {author} {\bibfnamefont {T.}~\bibnamefont
  {Vicsek}}\ and\ \bibinfo {author} {\bibfnamefont {A.}~\bibnamefont
  {Zafeiris}},\ }\Doi {10.1016/j.physrep.2012.03.004} {\bibfield  {journal}
  {\bibinfo  {journal} {Phys. Rep.},\ }\textbf {\bibinfo {volume} {517}},\
  \bibinfo {pages} {71} (\bibinfo {year} {2012})}\BibitemShut {NoStop}%
\bibitem [{\citenamefont {Helbing}\ \emph {et~al.}(2000)\citenamefont
  {Helbing}, \citenamefont {Farkas},\ and\ \citenamefont {Vicsek}}]{Helbing}%
  \BibitemOpen
  \bibfield  {author} {\bibinfo {author} {\bibfnamefont {D.}~\bibnamefont
  {Helbing}}, \bibinfo {author} {\bibfnamefont {I.}~\bibnamefont {Farkas}}, \
  and\ \bibinfo {author} {\bibfnamefont {T.}~\bibnamefont {Vicsek}},\ }\Doi
  {10.1038/35035023} {\bibfield  {journal} {\bibinfo  {journal} {Nature},\
  }\textbf {\bibinfo {volume} {407}},\ \bibinfo {pages} {487} (\bibinfo {year}
  {2000})}\BibitemShut {NoStop}%
\bibitem [{\citenamefont {Zhang}\ \emph {et~al.}(2012)\citenamefont {Zhang},
  \citenamefont {Klingsch}, \citenamefont {Schadschneider},\ and\ \citenamefont
  {Seyfried}}]{Schadschneider}%
  \BibitemOpen
  \bibfield  {author} {\bibinfo {author} {\bibfnamefont {J.}~\bibnamefont
  {Zhang}}, \bibinfo {author} {\bibfnamefont {W.}~\bibnamefont {Klingsch}},
  \bibinfo {author} {\bibfnamefont {A.}~\bibnamefont {Schadschneider}}, \ and\
  \bibinfo {author} {\bibfnamefont {A.}~\bibnamefont {Seyfried}},\ }\Doi
  {10.1088/1742-5468/2012/02/P02002} {\bibfield  {journal} {\bibinfo  {journal}
  {J. Stat Mech.},\ }\textbf {\bibinfo {volume} {2012}},\ \bibinfo {pages}
  {P02002} (\bibinfo {year} {2012})}\BibitemShut {NoStop}%
\bibitem [{\citenamefont {Silverberg}\ \emph {et~al.}(2013)\citenamefont
  {Silverberg}, \citenamefont {Bierbaum}, \citenamefont {Sethna},\ and\
  \citenamefont {Cohen}}]{Silverberg:13}%
  \BibitemOpen
  \bibfield  {author} {\bibinfo {author} {\bibfnamefont {J.~L.}\ \bibnamefont
  {Silverberg}}, \bibinfo {author} {\bibfnamefont {M.}~\bibnamefont
  {Bierbaum}}, \bibinfo {author} {\bibfnamefont {J.~P.}\ \bibnamefont
  {Sethna}}, \ and\ \bibinfo {author} {\bibfnamefont {I.}~\bibnamefont
  {Cohen}},\ }\Doi {10.1103/PhysRevLett.110.228701} {\bibfield  {journal}
  {\bibinfo  {journal} {Phys. Rev. Lett.},\ }\textbf {\bibinfo {volume}
  {110}},\ \bibinfo {pages} {228701} (\bibinfo {year} {2013})}\BibitemShut
  {NoStop}%
\bibitem [{\citenamefont {Dreyfus}\ \emph {et~al.}(2005)\citenamefont
  {Dreyfus}, \citenamefont {Baudry}, \citenamefont {Roper}, \citenamefont
  {Fermigier}, \citenamefont {Stone},\ and\ \citenamefont {Bibette}}]{Bibette}%
  \BibitemOpen
  \bibfield  {author} {\bibinfo {author} {\bibfnamefont {R.}~\bibnamefont
  {Dreyfus}}, \bibinfo {author} {\bibfnamefont {J.}~\bibnamefont {Baudry}},
  \bibinfo {author} {\bibfnamefont {M.~L.}\ \bibnamefont {Roper}}, \bibinfo
  {author} {\bibfnamefont {M.}~\bibnamefont {Fermigier}}, \bibinfo {author}
  {\bibfnamefont {H.~A.}\ \bibnamefont {Stone}}, \ and\ \bibinfo {author}
  {\bibfnamefont {J.}~\bibnamefont {Bibette}},\ }\Doi {10.1038/nature04090}
  {\bibfield  {journal} {\bibinfo  {journal} {Nature},\ }\textbf {\bibinfo
  {volume} {437}},\ \bibinfo {pages} {862} (\bibinfo {year}
  {2005})}\BibitemShut {NoStop}%
\bibitem [{\citenamefont {Thutupalli}\ \emph {et~al.}(2011)\citenamefont
  {Thutupalli}, \citenamefont {Seemann},\ and\ \citenamefont
  {Herminghaus}}]{2011Herminghaus}%
  \BibitemOpen
  \bibfield  {author} {\bibinfo {author} {\bibfnamefont {S.}~\bibnamefont
  {Thutupalli}}, \bibinfo {author} {\bibfnamefont {R.}~\bibnamefont {Seemann}},
  \ and\ \bibinfo {author} {\bibfnamefont {S.}~\bibnamefont {Herminghaus}},\
  }\Doi {10.1088/1367-2630/13/7/073021} {\bibfield  {journal} {\bibinfo
  {journal} {New J. Phys.},\ }\textbf {\bibinfo {volume} {13}},\ \bibinfo
  {pages} {073021} (\bibinfo {year} {2011})}\BibitemShut {NoStop}%
\bibitem [{\citenamefont {Reinm\"uller}\ \emph {et~al.}(2013)\citenamefont
  {Reinm\"uller}, \citenamefont {Sch\"ope},\ and\ \citenamefont
  {Palberg}}]{Reinmueller}%
  \BibitemOpen
  \bibfield  {author} {\bibinfo {author} {\bibfnamefont {A.}~\bibnamefont
  {Reinm\"uller}}, \bibinfo {author} {\bibfnamefont {H.~J.}\ \bibnamefont
  {Sch\"ope}}, \ and\ \bibinfo {author} {\bibfnamefont {T.}~\bibnamefont
  {Palberg}},\ }\href@noop {} {\bibfield  {journal} {\bibinfo  {journal}
  {Langmuir},\ }\textbf {\bibinfo {volume} {29}},\ \bibinfo {pages} {1738}
  (\bibinfo {year} {2013})}\BibitemShut {NoStop}%
\bibitem [{\citenamefont {Jiang}\ \emph {et~al.}(2010)\citenamefont {Jiang},
  \citenamefont {Yoshinaga},\ and\ \citenamefont {Sano}}]{Sano_PRL2010}%
  \BibitemOpen
  \bibfield  {author} {\bibinfo {author} {\bibfnamefont {H.-R.}\ \bibnamefont
  {Jiang}}, \bibinfo {author} {\bibfnamefont {N.}~\bibnamefont {Yoshinaga}}, \
  and\ \bibinfo {author} {\bibfnamefont {M.}~\bibnamefont {Sano}},\ }\Doi
  {10.1103/PhysRevLett.105.268302} {\bibfield  {journal} {\bibinfo  {journal}
  {Phys. Rev. Lett.},\ }\textbf {\bibinfo {volume} {105}},\ \bibinfo {pages}
  {268302} (\bibinfo {year} {2010})}\BibitemShut {NoStop}%
\bibitem [{\citenamefont {Buttinoni}\ \emph {et~al.}(2012)\citenamefont
  {Buttinoni}, \citenamefont {Volpe}, \citenamefont {K\"ummel}, \citenamefont
  {Volpe},\ and\ \citenamefont {Bechinger}}]{BechingerJPCM}%
  \BibitemOpen
  \bibfield  {author} {\bibinfo {author} {\bibfnamefont {I.}~\bibnamefont
  {Buttinoni}}, \bibinfo {author} {\bibfnamefont {G.}~\bibnamefont {Volpe}},
  \bibinfo {author} {\bibfnamefont {F.}~\bibnamefont {K\"ummel}}, \bibinfo
  {author} {\bibfnamefont {G.}~\bibnamefont {Volpe}}, \ and\ \bibinfo {author}
  {\bibfnamefont {C.}~\bibnamefont {Bechinger}},\ }\Doi
  {10.1088/0953-8984/24/28/284129} {\bibfield  {journal} {\bibinfo  {journal}
  {J. Phys.: Condens. Matter},\ }\textbf {\bibinfo {volume} {24}},\ \bibinfo
  {pages} {284129} (\bibinfo {year} {2012})}\BibitemShut {NoStop}%
\bibitem [{\citenamefont {Snezhko}\ and\ \citenamefont
  {Aranson}(2011)}]{snezhko_nature}%
  \BibitemOpen
  \bibfield  {author} {\bibinfo {author} {\bibfnamefont {A.}~\bibnamefont
  {Snezhko}}\ and\ \bibinfo {author} {\bibfnamefont {I.~S.}\ \bibnamefont
  {Aranson}},\ }\Doi {10.1038/NMAT3083} {\bibfield  {journal} {\bibinfo
  {journal} {Nature Mat.},\ }\textbf {\bibinfo {volume} {10}},\ \bibinfo
  {pages} {698} (\bibinfo {year} {2011})}\BibitemShut {NoStop}%
\bibitem [{\citenamefont {R\"uckner}\ and\ \citenamefont
  {Kapral}(2007)}]{2007Kapral}%
  \BibitemOpen
  \bibfield  {author} {\bibinfo {author} {\bibfnamefont {G.}~\bibnamefont
  {R\"uckner}}\ and\ \bibinfo {author} {\bibfnamefont {R.}~\bibnamefont
  {Kapral}},\ }\Doi {10.1103/PhysRevLett.98.150603} {\bibfield  {journal}
  {\bibinfo  {journal} {Phys. Rev. Lett.},\ }\textbf {\bibinfo {volume} {98}},\
  \bibinfo {pages} {150603} (\bibinfo {year} {2007})}\BibitemShut {NoStop}%
\bibitem [{\citenamefont {Tierno}\ \emph {et~al.}(2008)\citenamefont {Tierno},
  \citenamefont {Golestanian}, \citenamefont {Pagonabarraga},\ and\
  \citenamefont {Sagu\'{e}�s}}]{Sagues}%
  \BibitemOpen
  \bibfield  {author} {\bibinfo {author} {\bibfnamefont {P.}~\bibnamefont
  {Tierno}}, \bibinfo {author} {\bibfnamefont {R.}~\bibnamefont {Golestanian}},
  \bibinfo {author} {\bibfnamefont {I.}~\bibnamefont {Pagonabarraga}}, \ and\
  \bibinfo {author} {\bibfnamefont {F.}~\bibnamefont {Sagu\'{e}�s}},\ }\Doi
  {10.1021/jp808354n} {\bibfield  {journal} {\bibinfo  {journal} {J. Phys.
  Chem. B},\ }\textbf {\bibinfo {volume} {112}},\ \bibinfo {pages} {16525}
  (\bibinfo {year} {2008})}\BibitemShut {NoStop}%
\bibitem [{\citenamefont {Paxton}\ \emph {et~al.}(2005)\citenamefont {Paxton},
  \citenamefont {Sen},\ and\ \citenamefont {Mallouk}}]{Paxton}%
  \BibitemOpen
  \bibfield  {author} {\bibinfo {author} {\bibfnamefont {W.~F.}\ \bibnamefont
  {Paxton}}, \bibinfo {author} {\bibfnamefont {A.}~\bibnamefont {Sen}}, \ and\
  \bibinfo {author} {\bibfnamefont {T.~E.}\ \bibnamefont {Mallouk}},\
  }\href@noop {} {\bibfield  {journal} {\bibinfo  {journal} {Chem. Eur. J.},\
  }\textbf {\bibinfo {volume} {11}},\ \bibinfo {pages} {6462} (\bibinfo {year}
  {2005})}\BibitemShut {NoStop}%
\bibitem [{\citenamefont {Erbe}\ \emph {et~al.}(2008)\citenamefont {Erbe},
  \citenamefont {Zientara}, \citenamefont {Baraban}, \citenamefont {Kreidler},\
  and\ \citenamefont {Leiderer}}]{ErbeBaraban}%
  \BibitemOpen
  \bibfield  {author} {\bibinfo {author} {\bibfnamefont {A.}~\bibnamefont
  {Erbe}}, \bibinfo {author} {\bibfnamefont {M.}~\bibnamefont {Zientara}},
  \bibinfo {author} {\bibfnamefont {L.}~\bibnamefont {Baraban}}, \bibinfo
  {author} {\bibfnamefont {C.}~\bibnamefont {Kreidler}}, \ and\ \bibinfo
  {author} {\bibfnamefont {P.}~\bibnamefont {Leiderer}},\ }\Doi
  {10.1088/0953-8984/20/40/404215} {\bibfield  {journal} {\bibinfo  {journal}
  {J. Phys.: Condens. Matter},\ }\textbf {\bibinfo {volume} {20}},\ \bibinfo
  {pages} {404215} (\bibinfo {year} {2008})}\BibitemShut {NoStop}%
\bibitem [{\citenamefont {Anderson}(1989)}]{Anderson}%
  \BibitemOpen
  \bibfield  {author} {\bibinfo {author} {\bibfnamefont {J.~L.}\ \bibnamefont
  {Anderson}},\ }\href@noop {} {\bibfield  {journal} {\bibinfo  {journal}
  {Annu. Rev. Fluid Mech.},\ }\textbf {\bibinfo {volume} {21}},\ \bibinfo
  {pages} {61} (\bibinfo {year} {1989})}\BibitemShut {NoStop}%
\bibitem [{\citenamefont {Golestanian}\ \emph {et~al.}(2007)\citenamefont
  {Golestanian}, \citenamefont {Liverpool},\ and\ \citenamefont
  {Ajdari}}]{Golestanian_2007}%
  \BibitemOpen
  \bibfield  {author} {\bibinfo {author} {\bibfnamefont {R.}~\bibnamefont
  {Golestanian}}, \bibinfo {author} {\bibfnamefont {T.~B.}\ \bibnamefont
  {Liverpool}}, \ and\ \bibinfo {author} {\bibfnamefont {A.}~\bibnamefont
  {Ajdari}},\ }\href@noop {} {\bibfield  {journal} {\bibinfo  {journal} {New J.
  Phys.},\ }\textbf {\bibinfo {volume} {9}},\ \bibinfo {pages} {126} (\bibinfo
  {year} {2007})}\BibitemShut {NoStop}%
\bibitem [{\citenamefont {Ismagilov}\ \emph {et~al.}(2002)\citenamefont
  {Ismagilov}, \citenamefont {Schwartz}, \citenamefont {Bowden},\ and\
  \citenamefont {Whitesides}}]{Ismagilov}%
  \BibitemOpen
  \bibfield  {author} {\bibinfo {author} {\bibfnamefont {R.~F.}\ \bibnamefont
  {Ismagilov}}, \bibinfo {author} {\bibfnamefont {A.}~\bibnamefont {Schwartz}},
  \bibinfo {author} {\bibfnamefont {N.}~\bibnamefont {Bowden}}, \ and\ \bibinfo
  {author} {\bibfnamefont {G.~M.}\ \bibnamefont {Whitesides}},\ }\href@noop {}
  {\bibfield  {journal} {\bibinfo  {journal} {Angew. Chem. Int. Ed.},\ }\textbf
  {\bibinfo {volume} {41}},\ \bibinfo {pages} {652} (\bibinfo {year}
  {2002})}\BibitemShut {NoStop}%
\bibitem [{\citenamefont {Popescu}\ \emph {et~al.}(2009)\citenamefont
  {Popescu}, \citenamefont {Dietrich},\ and\ \citenamefont
  {Oshanin}}]{Dietrich}%
  \BibitemOpen
  \bibfield  {author} {\bibinfo {author} {\bibfnamefont {M.~N.}\ \bibnamefont
  {Popescu}}, \bibinfo {author} {\bibfnamefont {S.}~\bibnamefont {Dietrich}}, \
  and\ \bibinfo {author} {\bibfnamefont {G.}~\bibnamefont {Oshanin}},\ }\Doi
  {10.1063/1.3133239} {\bibfield  {journal} {\bibinfo  {journal} {J. Chem.
  Phys.},\ }\textbf {\bibinfo {volume} {130}},\ \bibinfo {pages} {194702}
  (\bibinfo {year} {2009})}\BibitemShut {NoStop}%
\bibitem [{\citenamefont {Ke}\ \emph {et~al.}(2010)\citenamefont {Ke},
  \citenamefont {Ye}, \citenamefont {Carroll},\ and\ \citenamefont
  {Showalter}}]{Ke}%
  \BibitemOpen
  \bibfield  {author} {\bibinfo {author} {\bibfnamefont {H.}~\bibnamefont
  {Ke}}, \bibinfo {author} {\bibfnamefont {S.-R.}\ \bibnamefont {Ye}}, \bibinfo
  {author} {\bibfnamefont {R.~L.}\ \bibnamefont {Carroll}}, \ and\ \bibinfo
  {author} {\bibfnamefont {K.}~\bibnamefont {Showalter}},\ }\href@noop {}
  {\bibfield  {journal} {\bibinfo  {journal} {J. Phys. Chem. A.},\ }\textbf
  {\bibinfo {volume} {114}},\ \bibinfo {pages} {5462} (\bibinfo {year}
  {2010})}\BibitemShut {NoStop}%
\bibitem [{\citenamefont {Ebbens}\ and\ \citenamefont
  {Howse}(2011)}]{Ebbens2011}%
  \BibitemOpen
  \bibfield  {author} {\bibinfo {author} {\bibfnamefont {S.~J.}\ \bibnamefont
  {Ebbens}}\ and\ \bibinfo {author} {\bibfnamefont {J.~R.}\ \bibnamefont
  {Howse}},\ }\href@noop {} {\bibfield  {journal} {\bibinfo  {journal}
  {Langmuir},\ }\textbf {\bibinfo {volume} {27}},\ \bibinfo {pages} {12293}
  (\bibinfo {year} {2011})}\BibitemShut {NoStop}%
\bibitem [{\citenamefont {Ebbens}\ \emph {et~al.}(2012)\citenamefont {Ebbens},
  \citenamefont {Tu}, \citenamefont {Howse},\ and\ \citenamefont
  {Golestanian}}]{Golestanian_2012}%
  \BibitemOpen
  \bibfield  {author} {\bibinfo {author} {\bibfnamefont {S.}~\bibnamefont
  {Ebbens}}, \bibinfo {author} {\bibfnamefont {M.-T.}\ \bibnamefont {Tu}},
  \bibinfo {author} {\bibfnamefont {J.~R.}\ \bibnamefont {Howse}}, \ and\
  \bibinfo {author} {\bibfnamefont {R.}~\bibnamefont {Golestanian}},\
  }\href@noop {} {\bibfield  {journal} {\bibinfo  {journal} {Phys. Rev. E},\
  }\textbf {\bibinfo {volume} {85}},\ \bibinfo {pages} {020401R(R)} (\bibinfo
  {year} {2012})}\BibitemShut {NoStop}%
\bibitem [{\citenamefont {Sabass}\ and\ \citenamefont
  {Seifert}(2012)}]{SabassJCP}%
  \BibitemOpen
  \bibfield  {author} {\bibinfo {author} {\bibfnamefont {B.}~\bibnamefont
  {Sabass}}\ and\ \bibinfo {author} {\bibfnamefont {U.}~\bibnamefont
  {Seifert}},\ }\href@noop {} {\bibfield  {journal} {\bibinfo  {journal} {J.\
  Chem.\ Phys.},\ }\textbf {\bibinfo {volume} {136}},\ \bibinfo {pages}
  {064508} (\bibinfo {year} {2012})}\BibitemShut {NoStop}%
\bibitem [{\citenamefont {Bickel}\ \emph {et~al.}(2013)\citenamefont {Bickel},
  \citenamefont {Majee},\ and\ \citenamefont {W\"urger}}]{Bickel_PRE2013}%
  \BibitemOpen
  \bibfield  {author} {\bibinfo {author} {\bibfnamefont {T.}~\bibnamefont
  {Bickel}}, \bibinfo {author} {\bibfnamefont {A.}~\bibnamefont {Majee}}, \
  and\ \bibinfo {author} {\bibfnamefont {A.}~\bibnamefont {W\"urger}},\ }\Doi
  {10.1103/PhysRevE.88.012301} {\bibfield  {journal} {\bibinfo  {journal}
  {Phys. Rev. E},\ }\textbf {\bibinfo {volume} {88}},\ \bibinfo {pages}
  {012301} (\bibinfo {year} {2013})}\BibitemShut {NoStop}%
\bibitem [{\citenamefont {Buttinoni}\ \emph {et~al.}(2013)\citenamefont
  {Buttinoni}, \citenamefont {Bialk\'e}, \citenamefont {K\"ummel},
  \citenamefont {L\"owen}, \citenamefont {Bechinger},\ and\ \citenamefont
  {Speck}}]{Bialke_PRL2013}%
  \BibitemOpen
  \bibfield  {author} {\bibinfo {author} {\bibfnamefont {I.}~\bibnamefont
  {Buttinoni}}, \bibinfo {author} {\bibfnamefont {J.}~\bibnamefont {Bialk\'e}},
  \bibinfo {author} {\bibfnamefont {F.}~\bibnamefont {K\"ummel}}, \bibinfo
  {author} {\bibfnamefont {H.}~\bibnamefont {L\"owen}}, \bibinfo {author}
  {\bibfnamefont {C.}~\bibnamefont {Bechinger}}, \ and\ \bibinfo {author}
  {\bibfnamefont {T.}~\bibnamefont {Speck}},\ }\Doi
  {10.1103/PhysRevLett.110.238301} {\bibfield  {journal} {\bibinfo  {journal}
  {Phys. Rev. Lett.},\ }\textbf {\bibinfo {volume} {110}},\ \bibinfo {pages}
  {238301} (\bibinfo {year} {2013})}\BibitemShut {NoStop}%
\bibitem [{\citenamefont {Theurkauff}\ \emph {et~al.}(2012)\citenamefont
  {Theurkauff}, \citenamefont {Cottin-Bizonne}, \citenamefont {Palacci},
  \citenamefont {Ybert},\ and\ \citenamefont
  {Bocquet}}]{theurkauff2012dynamic}%
  \BibitemOpen
  \bibfield  {author} {\bibinfo {author} {\bibfnamefont {I.}~\bibnamefont
  {Theurkauff}}, \bibinfo {author} {\bibfnamefont {C.}~\bibnamefont
  {Cottin-Bizonne}}, \bibinfo {author} {\bibfnamefont {J.}~\bibnamefont
  {Palacci}}, \bibinfo {author} {\bibfnamefont {C.}~\bibnamefont {Ybert}}, \
  and\ \bibinfo {author} {\bibfnamefont {L.}~\bibnamefont {Bocquet}},\
  }\href@noop {} {\bibfield  {journal} {\bibinfo  {journal} {Phys. Rev.
  Lett.},\ }\textbf {\bibinfo {volume} {108}},\ \bibinfo {pages} {268303}
  (\bibinfo {year} {2012})}\BibitemShut {NoStop}%
\bibitem [{\citenamefont {Palacci}\ \emph {et~al.}(2013)\citenamefont
  {Palacci}, \citenamefont {Sacanna}, \citenamefont {Steinberg}, \citenamefont
  {Pine},\ and\ \citenamefont {Chaikin}}]{ChaikinScience}%
  \BibitemOpen
  \bibfield  {author} {\bibinfo {author} {\bibfnamefont {J.}~\bibnamefont
  {Palacci}}, \bibinfo {author} {\bibfnamefont {S.}~\bibnamefont {Sacanna}},
  \bibinfo {author} {\bibfnamefont {A.~P.}\ \bibnamefont {Steinberg}}, \bibinfo
  {author} {\bibfnamefont {D.~J.}\ \bibnamefont {Pine}}, \ and\ \bibinfo
  {author} {\bibfnamefont {P.~M.}\ \bibnamefont {Chaikin}},\ }\href@noop {}
  {\bibfield  {journal} {\bibinfo  {journal} {Science},\ }\textbf {\bibinfo
  {volume} {339}},\ \bibinfo {pages} {936} (\bibinfo {year}
  {2013})}\BibitemShut {NoStop}%
\bibitem [{\citenamefont {Redner}\ \emph {et~al.}(2013)\citenamefont {Redner},
  \citenamefont {Hagan},\ and\ \citenamefont {Baskaran}}]{Baskaran_PRL2013}%
  \BibitemOpen
  \bibfield  {author} {\bibinfo {author} {\bibfnamefont {G.~S.}\ \bibnamefont
  {Redner}}, \bibinfo {author} {\bibfnamefont {M.~F.}\ \bibnamefont {Hagan}}, \
  and\ \bibinfo {author} {\bibfnamefont {A.}~\bibnamefont {Baskaran}},\ }\Doi
  {10.1103/PhysRevLett.110.055701} {\bibfield  {journal} {\bibinfo  {journal}
  {Phys. Rev. Lett.},\ }\textbf {\bibinfo {volume} {110}},\ \bibinfo {pages}
  {055701} (\bibinfo {year} {2013})}\BibitemShut {NoStop}%
\bibitem [{\citenamefont {Baraban}\ \emph
  {et~al.}(2012){\natexlab{a}}\citenamefont {Baraban}, \citenamefont
  {Tasinkevych}, \citenamefont {Popescu}, \citenamefont {Sanchez},
  \citenamefont {Dietrich},\ and\ \citenamefont
  {Schmidt}}]{Baraban_SoftMatter}%
  \BibitemOpen
  \bibfield  {author} {\bibinfo {author} {\bibfnamefont {L.}~\bibnamefont
  {Baraban}}, \bibinfo {author} {\bibfnamefont {M.}~\bibnamefont
  {Tasinkevych}}, \bibinfo {author} {\bibfnamefont {M.~N.}\ \bibnamefont
  {Popescu}}, \bibinfo {author} {\bibfnamefont {S.}~\bibnamefont {Sanchez}},
  \bibinfo {author} {\bibfnamefont {S.}~\bibnamefont {Dietrich}}, \ and\
  \bibinfo {author} {\bibfnamefont {O.~G.}\ \bibnamefont {Schmidt}},\
  }\href@noop {} {\bibfield  {journal} {\bibinfo  {journal} {Soft Matter},\
  }\textbf {\bibinfo {volume} {8}},\ \bibinfo {pages} {48} (\bibinfo {year}
  {2012}{\natexlab{a}})}\BibitemShut {NoStop}%
\bibitem [{\citenamefont {Baraban}\ \emph
  {et~al.}(2012){\natexlab{b}}\citenamefont {Baraban}, \citenamefont {Makarov},
  \citenamefont {Streubel}, \citenamefont {M\"onch}, \citenamefont {Grimm},
  \citenamefont {Sanchez},\ and\ \citenamefont {Schmidt}}]{Baraban_ACSnano}%
  \BibitemOpen
  \bibfield  {author} {\bibinfo {author} {\bibfnamefont {L.}~\bibnamefont
  {Baraban}}, \bibinfo {author} {\bibfnamefont {D.}~\bibnamefont {Makarov}},
  \bibinfo {author} {\bibfnamefont {R.}~\bibnamefont {Streubel}}, \bibinfo
  {author} {\bibfnamefont {I.}~\bibnamefont {M\"onch}}, \bibinfo {author}
  {\bibfnamefont {D.}~\bibnamefont {Grimm}}, \bibinfo {author} {\bibfnamefont
  {S.}~\bibnamefont {Sanchez}}, \ and\ \bibinfo {author} {\bibfnamefont
  {O.~G.}\ \bibnamefont {Schmidt}},\ }\href@noop {} {\bibfield  {journal}
  {\bibinfo  {journal} {ACS Nano},\ }\textbf {\bibinfo {volume} {6}},\ \bibinfo
  {pages} {3383} (\bibinfo {year} {2012}{\natexlab{b}})}\BibitemShut {NoStop}%
\bibitem [{\citenamefont {Baraban}\ \emph {et~al.}(2013)\citenamefont
  {Baraban}, \citenamefont {Makarov}, \citenamefont {Schmidt}, \citenamefont
  {Cuniberti}, \citenamefont {Leiderer},\ and\ \citenamefont
  {Erbe}}]{Baraban_2013}%
  \BibitemOpen
  \bibfield  {author} {\bibinfo {author} {\bibfnamefont {L.}~\bibnamefont
  {Baraban}}, \bibinfo {author} {\bibfnamefont {D.}~\bibnamefont {Makarov}},
  \bibinfo {author} {\bibfnamefont {O.}~\bibnamefont {Schmidt}}, \bibinfo
  {author} {\bibfnamefont {G.}~\bibnamefont {Cuniberti}}, \bibinfo {author}
  {\bibfnamefont {P.}~\bibnamefont {Leiderer}}, \ and\ \bibinfo {author}
  {\bibfnamefont {A.}~\bibnamefont {Erbe}},\ }\Doi {10.1039/c2nr32662k}
  {\bibfield  {journal} {\bibinfo  {journal} {Nanoscale},\ }\textbf {\bibinfo
  {volume} {5}},\ \bibinfo {pages} {1332} (\bibinfo {year} {2013})}\BibitemShut
  {NoStop}%
\bibitem [{\citenamefont {Volpe}\ \emph {et~al.}(2011)\citenamefont {Volpe},
  \citenamefont {Buttinoni}, \citenamefont {Vogt}, \citenamefont {K\"ummerer},\
  and\ \citenamefont {Bechinger}}]{VolpeSM1}%
  \BibitemOpen
  \bibfield  {author} {\bibinfo {author} {\bibfnamefont {G.}~\bibnamefont
  {Volpe}}, \bibinfo {author} {\bibfnamefont {I.}~\bibnamefont {Buttinoni}},
  \bibinfo {author} {\bibfnamefont {D.}~\bibnamefont {Vogt}}, \bibinfo {author}
  {\bibfnamefont {H.-J.}\ \bibnamefont {K\"ummerer}}, \ and\ \bibinfo {author}
  {\bibfnamefont {C.}~\bibnamefont {Bechinger}},\ }\Doi {10.1039/C1SM05960B}
  {\bibfield  {journal} {\bibinfo  {journal} {Soft Matter},\ }\textbf {\bibinfo
  {volume} {7}},\ \bibinfo {pages} {8810} (\bibinfo {year} {2011})}\BibitemShut
  {NoStop}%
\bibitem [{\citenamefont {Mijalkov}\ and\ \citenamefont
  {Volpe}(2013)}]{VolpeSorting}%
  \BibitemOpen
  \bibfield  {author} {\bibinfo {author} {\bibfnamefont {M.}~\bibnamefont
  {Mijalkov}}\ and\ \bibinfo {author} {\bibfnamefont {G.}~\bibnamefont
  {Volpe}},\ }\href@noop {} {\bibfield  {journal} {\bibinfo  {journal} {Soft
  Matter},\ }\textbf {\bibinfo {volume} {9}},\ \bibinfo {pages} {6376}
  (\bibinfo {year} {2013})}\BibitemShut {NoStop}%
\bibitem [{\citenamefont {Yang}\ \emph {et~al.}(2012)\citenamefont {Yang},
  \citenamefont {Misko}, \citenamefont {Nelissen}, \citenamefont {Kong},\ and\
  \citenamefont {Peeters}}]{YangSeparation}%
  \BibitemOpen
  \bibfield  {author} {\bibinfo {author} {\bibfnamefont {W.}~\bibnamefont
  {Yang}}, \bibinfo {author} {\bibfnamefont {V.~R.}\ \bibnamefont {Misko}},
  \bibinfo {author} {\bibfnamefont {K.}~\bibnamefont {Nelissen}}, \bibinfo
  {author} {\bibfnamefont {M.}~\bibnamefont {Kong}}, \ and\ \bibinfo {author}
  {\bibfnamefont {F.~M.}\ \bibnamefont {Peeters}},\ }\href@noop {} {\bibfield
  {journal} {\bibinfo  {journal} {Soft Matter},\ }\textbf {\bibinfo {volume}
  {8}},\ \bibinfo {pages} {5175} (\bibinfo {year} {2012})}\BibitemShut
  {NoStop}%
\bibitem [{\citenamefont {Ghosh}\ \emph {et~al.}(2013)\citenamefont {Ghosh},
  \citenamefont {Misko}, \citenamefont {Marchesoni},\ and\ \citenamefont
  {Nori}}]{Ghoshratchet}%
  \BibitemOpen
  \bibfield  {author} {\bibinfo {author} {\bibfnamefont {P.~K.}\ \bibnamefont
  {Ghosh}}, \bibinfo {author} {\bibfnamefont {V.~R.}\ \bibnamefont {Misko}},
  \bibinfo {author} {\bibfnamefont {F.}~\bibnamefont {Marchesoni}}, \ and\
  \bibinfo {author} {\bibfnamefont {F.}~\bibnamefont {Nori}},\ }\href@noop {}
  {\bibfield  {journal} {\bibinfo  {journal} {Phys. Rev. Lett.},\ }\textbf
  {\bibinfo {volume} {110}},\ \bibinfo {pages} {268301} (\bibinfo {year}
  {2013})}\BibitemShut {NoStop}%
\bibitem [{\citenamefont {Howse}\ \emph {et~al.}(2007)\citenamefont {Howse},
  \citenamefont {Jones}, \citenamefont {Ryan}, \citenamefont {Gough},
  \citenamefont {Vafabakhsh},\ and\ \citenamefont {Golestanian}}]{Howse_2007}%
  \BibitemOpen
  \bibfield  {author} {\bibinfo {author} {\bibfnamefont {J.~R.}\ \bibnamefont
  {Howse}}, \bibinfo {author} {\bibfnamefont {R.~A.~L.}\ \bibnamefont {Jones}},
  \bibinfo {author} {\bibfnamefont {A.~J.}\ \bibnamefont {Ryan}}, \bibinfo
  {author} {\bibfnamefont {T.}~\bibnamefont {Gough}}, \bibinfo {author}
  {\bibfnamefont {R.}~\bibnamefont {Vafabakhsh}}, \ and\ \bibinfo {author}
  {\bibfnamefont {R.}~\bibnamefont {Golestanian}},\ }\href@noop {} {\bibfield
  {journal} {\bibinfo  {journal} {Phys. Rev. Lett.},\ }\textbf {\bibinfo
  {volume} {99}},\ \bibinfo {pages} {048102} (\bibinfo {year}
  {2007})}\BibitemShut {NoStop}%
\bibitem [{\citenamefont {ten Hagen}\ \emph {et~al.}(2009)\citenamefont {ten
  Hagen}, \citenamefont {van Teeffelen},\ and\ \citenamefont
  {L\"owen}}]{Cond_Matt}%
  \BibitemOpen
  \bibfield  {author} {\bibinfo {author} {\bibfnamefont {B.}~\bibnamefont {ten
  Hagen}}, \bibinfo {author} {\bibfnamefont {S.}~\bibnamefont {van Teeffelen}},
  \ and\ \bibinfo {author} {\bibfnamefont {H.}~\bibnamefont {L\"owen}},\
  }\href@noop {} {\bibfield  {journal} {\bibinfo  {journal}
  {Condens.~Matter~Phys.},\ }\textbf {\bibinfo {volume} {12}},\ \bibinfo
  {pages} {725} (\bibinfo {year} {2009})}\BibitemShut {NoStop}%
\bibitem [{\citenamefont {ten Hagen}\ \emph
  {et~al.}(2011){\natexlab{a}}\citenamefont {ten Hagen}, \citenamefont {van
  Teeffelen},\ and\ \citenamefont {L\"owen}}]{tenHagen_JPCM}%
  \BibitemOpen
  \bibfield  {author} {\bibinfo {author} {\bibfnamefont {B.}~\bibnamefont {ten
  Hagen}}, \bibinfo {author} {\bibfnamefont {S.}~\bibnamefont {van Teeffelen}},
  \ and\ \bibinfo {author} {\bibfnamefont {H.}~\bibnamefont {L\"owen}},\
  }\href@noop {} {\bibfield  {journal} {\bibinfo  {journal} {J. Phys.: Condens.
  Matter},\ }\textbf {\bibinfo {volume} {23}},\ \bibinfo {pages} {194119}
  (\bibinfo {year} {2011}{\natexlab{a}})}\BibitemShut {NoStop}%
\bibitem [{\citenamefont {Kob}\ \emph {et~al.}(1997)\citenamefont {Kob},
  \citenamefont {Donati}, \citenamefont {Plimpton}, \citenamefont {Poole},\
  and\ \citenamefont {Glotzer}}]{KobPRL}%
  \BibitemOpen
  \bibfield  {author} {\bibinfo {author} {\bibfnamefont {W.}~\bibnamefont
  {Kob}}, \bibinfo {author} {\bibfnamefont {C.}~\bibnamefont {Donati}},
  \bibinfo {author} {\bibfnamefont {S.~J.}\ \bibnamefont {Plimpton}}, \bibinfo
  {author} {\bibfnamefont {P.~H.}\ \bibnamefont {Poole}}, \ and\ \bibinfo
  {author} {\bibfnamefont {S.~C.}\ \bibnamefont {Glotzer}},\ }\href@noop {}
  {\bibfield  {journal} {\bibinfo  {journal} {Phys.\ Rev.\ Lett.},\ }\textbf
  {\bibinfo {volume} {79}},\ \bibinfo {pages} {2827} (\bibinfo {year}
  {1997})}\BibitemShut {NoStop}%
\bibitem [{\citenamefont {Puertas}\ \emph {et~al.}(2003)\citenamefont
  {Puertas}, \citenamefont {Fuchs},\ and\ \citenamefont
  {Cates}}]{PuertasCates}%
  \BibitemOpen
  \bibfield  {author} {\bibinfo {author} {\bibfnamefont {A.~M.}\ \bibnamefont
  {Puertas}}, \bibinfo {author} {\bibfnamefont {M.}~\bibnamefont {Fuchs}}, \
  and\ \bibinfo {author} {\bibfnamefont {M.~E.}\ \bibnamefont {Cates}},\
  }\href@noop {} {\bibfield  {journal} {\bibinfo  {journal} {Phys.~Rev.~E},\
  }\textbf {\bibinfo {volume} {67}},\ \bibinfo {pages} {031406} (\bibinfo
  {year} {2003})}\BibitemShut {NoStop}%
\bibitem [{\citenamefont {Vollmayr-Lee}\ \emph {et~al.}(2002)\citenamefont
  {Vollmayr-Lee}, \citenamefont {Kob}, \citenamefont {Binder},\ and\
  \citenamefont {Zippelius}}]{BinderZippelius}%
  \BibitemOpen
  \bibfield  {author} {\bibinfo {author} {\bibfnamefont {K.}~\bibnamefont
  {Vollmayr-Lee}}, \bibinfo {author} {\bibfnamefont {W.}~\bibnamefont {Kob}},
  \bibinfo {author} {\bibfnamefont {K.}~\bibnamefont {Binder}}, \ and\ \bibinfo
  {author} {\bibfnamefont {A.}~\bibnamefont {Zippelius}},\ }\href@noop {}
  {\bibfield  {journal} {\bibinfo  {journal} {J.\ Chem.\ Phys.},\ }\textbf
  {\bibinfo {volume} {116}},\ \bibinfo {pages} {5158} (\bibinfo {year}
  {2002})}\BibitemShut {NoStop}%
\bibitem [{\citenamefont {Arbe}\ \emph {et~al.}(2002)\citenamefont {Arbe},
  \citenamefont {Colmenero}, \citenamefont {Alvarez}, \citenamefont
  {Monkenbusch}, \citenamefont {Richter}, \citenamefont {Farago},\ and\
  \citenamefont {Frick}}]{Arbe02}%
  \BibitemOpen
  \bibfield  {author} {\bibinfo {author} {\bibfnamefont {A.}~\bibnamefont
  {Arbe}}, \bibinfo {author} {\bibfnamefont {J.}~\bibnamefont {Colmenero}},
  \bibinfo {author} {\bibfnamefont {F.}~\bibnamefont {Alvarez}}, \bibinfo
  {author} {\bibfnamefont {M.}~\bibnamefont {Monkenbusch}}, \bibinfo {author}
  {\bibfnamefont {D.}~\bibnamefont {Richter}}, \bibinfo {author} {\bibfnamefont
  {B.}~\bibnamefont {Farago}}, \ and\ \bibinfo {author} {\bibfnamefont
  {B.}~\bibnamefont {Frick}},\ }\href@noop {} {\bibfield  {journal} {\bibinfo
  {journal} {Phys. Rev. Lett.},\ }\textbf {\bibinfo {volume} {89}},\ \bibinfo
  {pages} {245701} (\bibinfo {year} {2002})}\BibitemShut {NoStop}%
\bibitem [{\citenamefont {Bouchaud}\ and\ \citenamefont
  {Potters}(2009)}]{bouchaud_book}%
  \BibitemOpen
  \bibfield  {author} {\bibinfo {author} {\bibfnamefont {J.-P.}\ \bibnamefont
  {Bouchaud}}\ and\ \bibinfo {author} {\bibfnamefont {M.}~\bibnamefont
  {Potters}},\ }\href@noop {} {\emph {\bibinfo {title} {Theory of Financial
  Risk and Derivative Pricing: From Statistical Physics to Risk Management}}},\
  \bibinfo {edition} {2nd}\ ed.\ (\bibinfo  {publisher} {Cambridge University
  Press},\ \bibinfo {address} {Cambridge},\ \bibinfo {year} {2009})\BibitemShut
  {NoStop}%
\bibitem [{\citenamefont {Happel}\ and\ \citenamefont
  {Brenner}(1991)}]{HappelB1991}%
  \BibitemOpen
  \bibfield  {author} {\bibinfo {author} {\bibfnamefont {J.}~\bibnamefont
  {Happel}}\ and\ \bibinfo {author} {\bibfnamefont {H.}~\bibnamefont
  {Brenner}},\ }\href@noop {} {\emph {\bibinfo {title} {Low {R}eynolds Number
  Hydrodynamics: With Special Applications to Particulate Media}}},\ \bibinfo
  {edition} {2nd}\ ed.,\ \bibinfo {series} {Mechanics of Fluids and Transport
  Processes}, Vol.~\bibinfo {volume} {1}\ (\bibinfo  {publisher} {Kluwer
  Academic Publishers},\ \bibinfo {address} {Dordrecht},\ \bibinfo {year}
  {1991})\BibitemShut {NoStop}%
\bibitem [{\citenamefont {Jeffrey}(1992)}]{Jeffrey1992}%
  \BibitemOpen
  \bibfield  {author} {\bibinfo {author} {\bibfnamefont {D.~J.}\ \bibnamefont
  {Jeffrey}},\ }\href@noop {} {\bibfield  {journal} {\bibinfo  {journal} {Phys.
  Fluids A},\ }\textbf {\bibinfo {volume} {4}},\ \bibinfo {pages} {16}
  (\bibinfo {year} {1992})}\BibitemShut {NoStop}%
\bibitem [{\citenamefont {Dhont}(1996)}]{Dhont_book}%
  \BibitemOpen
  \bibfield  {author} {\bibinfo {author} {\bibfnamefont {J.~K.~G.}\
  \bibnamefont {Dhont}},\ }\href@noop {} {\emph {\bibinfo {title} {An
  Introduction to Dynamics of Colloids}}}\ (\bibinfo  {publisher} {Elsevier},\
  \bibinfo {address} {Amsterdam},\ \bibinfo {year} {1996})\BibitemShut
  {NoStop}%
\bibitem [{\citenamefont {K\"ummel}\ \emph {et~al.}(2013)\citenamefont
  {K\"ummel}, \citenamefont {ten Hagen}, \citenamefont {Wittkowski},
  \citenamefont {Buttinoni}, \citenamefont {Eichhorn}, \citenamefont {Volpe},
  \citenamefont {L\"owen},\ and\ \citenamefont {Bechinger}}]{Kuemmel:13}%
  \BibitemOpen
  \bibfield  {author} {\bibinfo {author} {\bibfnamefont {F.}~\bibnamefont
  {K\"ummel}}, \bibinfo {author} {\bibfnamefont {B.}~\bibnamefont {ten Hagen}},
  \bibinfo {author} {\bibfnamefont {R.}~\bibnamefont {Wittkowski}}, \bibinfo
  {author} {\bibfnamefont {I.}~\bibnamefont {Buttinoni}}, \bibinfo {author}
  {\bibfnamefont {R.}~\bibnamefont {Eichhorn}}, \bibinfo {author}
  {\bibfnamefont {G.}~\bibnamefont {Volpe}}, \bibinfo {author} {\bibfnamefont
  {H.}~\bibnamefont {L\"owen}}, \ and\ \bibinfo {author} {\bibfnamefont
  {C.}~\bibnamefont {Bechinger}},\ }\href@noop {} {\bibfield  {journal}
  {\bibinfo  {journal} {Phys. Rev. Lett.},\ }\textbf {\bibinfo {volume}
  {110}},\ \bibinfo {pages} {198302} (\bibinfo {year} {2013})}\BibitemShut
  {NoStop}%
\bibitem [{\citenamefont {Han}\ \emph {et~al.}(2006)\citenamefont {Han},
  \citenamefont {Alsayed}, \citenamefont {Nobili}, \citenamefont {Zhang},
  \citenamefont {Lubensky},\ and\ \citenamefont {Yodh}}]{Han:06}%
  \BibitemOpen
  \bibfield  {author} {\bibinfo {author} {\bibfnamefont {Y.}~\bibnamefont
  {Han}}, \bibinfo {author} {\bibfnamefont {A.~M.}\ \bibnamefont {Alsayed}},
  \bibinfo {author} {\bibfnamefont {M.}~\bibnamefont {Nobili}}, \bibinfo
  {author} {\bibfnamefont {J.}~\bibnamefont {Zhang}}, \bibinfo {author}
  {\bibfnamefont {T.~C.}\ \bibnamefont {Lubensky}}, \ and\ \bibinfo {author}
  {\bibfnamefont {A.~G.}\ \bibnamefont {Yodh}},\ }\href@noop {} {\bibfield
  {journal} {\bibinfo  {journal} {Science},\ }\textbf {\bibinfo {volume}
  {314}},\ \bibinfo {pages} {626} (\bibinfo {year} {2006})}\BibitemShut
  {NoStop}%
\bibitem [{\citenamefont {Marsaglia}(1972)}]{Marsaglia}%
  \BibitemOpen
  \bibfield  {author} {\bibinfo {author} {\bibfnamefont {G.}~\bibnamefont
  {Marsaglia}},\ }\href@noop {} {\bibfield  {journal} {\bibinfo  {journal}
  {Ann. Math. Stat.},\ }\textbf {\bibinfo {volume} {43}},\ \bibinfo {pages}
  {645} (\bibinfo {year} {1972})}\BibitemShut {NoStop}%
\bibitem [{\citenamefont {Gauger}\ and\ \citenamefont
  {Stark}(2006)}]{Gauger:06}%
  \BibitemOpen
  \bibfield  {author} {\bibinfo {author} {\bibfnamefont {E.}~\bibnamefont
  {Gauger}}\ and\ \bibinfo {author} {\bibfnamefont {H.}~\bibnamefont {Stark}},\
  }\href@noop {} {\bibfield  {journal} {\bibinfo  {journal} {Phys. Rev. E},\
  }\textbf {\bibinfo {volume} {74}},\ \bibinfo {pages} {021907} (\bibinfo
  {year} {2006})}\BibitemShut {NoStop}%
\bibitem [{\citenamefont {G\"otze}\ and\ \citenamefont
  {Gompper}(2010)}]{Goetze:10}%
  \BibitemOpen
  \bibfield  {author} {\bibinfo {author} {\bibfnamefont {I.~O.}\ \bibnamefont
  {G\"otze}}\ and\ \bibinfo {author} {\bibfnamefont {G.}~\bibnamefont
  {Gompper}},\ }\href@noop {} {\bibfield  {journal} {\bibinfo  {journal} {Phys.
  Rev. E},\ }\textbf {\bibinfo {volume} {82}},\ \bibinfo {pages} {041921}
  (\bibinfo {year} {2010})}\BibitemShut {NoStop}%
\bibitem [{\citenamefont {H\"ofling}\ \emph {et~al.}(2008)\citenamefont
  {H\"ofling}, \citenamefont {Frey},\ and\ \citenamefont
  {Franosch}}]{HoeflingPRL}%
  \BibitemOpen
  \bibfield  {author} {\bibinfo {author} {\bibfnamefont {F.}~\bibnamefont
  {H\"ofling}}, \bibinfo {author} {\bibfnamefont {E.}~\bibnamefont {Frey}}, \
  and\ \bibinfo {author} {\bibfnamefont {T.}~\bibnamefont {Franosch}},\
  }\href@noop {} {\bibfield  {journal} {\bibinfo  {journal} {Phys. Rev.
  Lett.},\ }\textbf {\bibinfo {volume} {101}},\ \bibinfo {pages} {120605}
  (\bibinfo {year} {2008})}\BibitemShut {NoStop}%
\bibitem [{\citenamefont {Han}\ \emph {et~al.}(2009)\citenamefont {Han},
  \citenamefont {Alsayed}, \citenamefont {Nobili},\ and\ \citenamefont
  {Yodh}}]{Han:09}%
  \BibitemOpen
  \bibfield  {author} {\bibinfo {author} {\bibfnamefont {Y.}~\bibnamefont
  {Han}}, \bibinfo {author} {\bibfnamefont {A.}~\bibnamefont {Alsayed}},
  \bibinfo {author} {\bibfnamefont {M.}~\bibnamefont {Nobili}}, \ and\ \bibinfo
  {author} {\bibfnamefont {A.~G.}\ \bibnamefont {Yodh}},\ }\href@noop {}
  {\bibfield  {journal} {\bibinfo  {journal} {Phys.~Rev.~E},\ }\textbf
  {\bibinfo {volume} {80}},\ \bibinfo {pages} {011403} (\bibinfo {year}
  {2009})}\BibitemShut {NoStop}%
\bibitem [{\citenamefont {Wittkowski}\ and\ \citenamefont
  {L{\"o}wen}(2012)}]{WittkowskiL2012}%
  \BibitemOpen
  \bibfield  {author} {\bibinfo {author} {\bibfnamefont {R.}~\bibnamefont
  {Wittkowski}}\ and\ \bibinfo {author} {\bibfnamefont {H.}~\bibnamefont
  {L{\"o}wen}},\ }\href@noop {} {\bibfield  {journal} {\bibinfo  {journal}
  {Phys.~Rev.~E},\ }\textbf {\bibinfo {volume} {85}},\ \bibinfo {pages}
  {021406} (\bibinfo {year} {2012})}\BibitemShut {NoStop}%
\bibitem [{\citenamefont {Kraft}\ \emph {et~al.}(2013)\citenamefont {Kraft},
  \citenamefont {Wittkowski}, \citenamefont {ten Hagen}, \citenamefont
  {Edmond}, \citenamefont {Pine},\ and\ \citenamefont {L{\"o}wen}}]{Kraft:13}%
  \BibitemOpen
  \bibfield  {author} {\bibinfo {author} {\bibfnamefont {D.~J.}\ \bibnamefont
  {Kraft}}, \bibinfo {author} {\bibfnamefont {R.}~\bibnamefont {Wittkowski}},
  \bibinfo {author} {\bibfnamefont {B.}~\bibnamefont {ten Hagen}}, \bibinfo
  {author} {\bibfnamefont {K.~V.}\ \bibnamefont {Edmond}}, \bibinfo {author}
  {\bibfnamefont {D.~J.}\ \bibnamefont {Pine}}, \ and\ \bibinfo {author}
  {\bibfnamefont {H.}~\bibnamefont {L{\"o}wen}},\ }\href@noop {} {\bibfield
  {journal} {\bibinfo  {journal} {arXiv preprint, arXiv:1305.1253}} (\bibinfo
  {year} {2013})}\BibitemShut {NoStop}%
\bibitem [{\citenamefont {van Teeffelen}\ and\ \citenamefont
  {L{\"o}wen}(2008)}]{vanTeeffelenL2008}%
  \BibitemOpen
  \bibfield  {author} {\bibinfo {author} {\bibfnamefont {S.}~\bibnamefont {van
  Teeffelen}}\ and\ \bibinfo {author} {\bibfnamefont {H.}~\bibnamefont
  {L{\"o}wen}},\ }\href@noop {} {\bibfield  {journal} {\bibinfo  {journal}
  {Phys. Rev. E},\ }\textbf {\bibinfo {volume} {78}},\ \bibinfo {pages}
  {020101(R)} (\bibinfo {year} {2008})}\BibitemShut {NoStop}%
\bibitem [{\citenamefont {Holzer}\ \emph {et~al.}(2010)\citenamefont {Holzer},
  \citenamefont {Bammert}, \citenamefont {Rzehak},\ and\ \citenamefont
  {Zimmermann}}]{Holzer:10}%
  \BibitemOpen
  \bibfield  {author} {\bibinfo {author} {\bibfnamefont {L.}~\bibnamefont
  {Holzer}}, \bibinfo {author} {\bibfnamefont {J.}~\bibnamefont {Bammert}},
  \bibinfo {author} {\bibfnamefont {R.}~\bibnamefont {Rzehak}}, \ and\ \bibinfo
  {author} {\bibfnamefont {W.}~\bibnamefont {Zimmermann}},\ }\href@noop {}
  {\bibfield  {journal} {\bibinfo  {journal} {Phys. Rev. E},\ }\textbf
  {\bibinfo {volume} {81}},\ \bibinfo {pages} {041124} (\bibinfo {year}
  {2010})}\BibitemShut {NoStop}%
\bibitem [{\citenamefont {Z\"ottl}\ and\ \citenamefont
  {Stark}(2012)}]{Zoettl:12}%
  \BibitemOpen
  \bibfield  {author} {\bibinfo {author} {\bibfnamefont {A.}~\bibnamefont
  {Z\"ottl}}\ and\ \bibinfo {author} {\bibfnamefont {H.}~\bibnamefont
  {Stark}},\ }\href@noop {} {\bibfield  {journal} {\bibinfo  {journal} {Phys.
  Rev. Lett.},\ }\textbf {\bibinfo {volume} {108}},\ \bibinfo {pages} {218104}
  (\bibinfo {year} {2012})}\BibitemShut {NoStop}%
\bibitem [{\citenamefont {ten Hagen}\ \emph
  {et~al.}(2011){\natexlab{b}}\citenamefont {ten Hagen}, \citenamefont
  {Wittkowski},\ and\ \citenamefont {L\"owen}}]{tenHagenPRE:11}%
  \BibitemOpen
  \bibfield  {author} {\bibinfo {author} {\bibfnamefont {B.}~\bibnamefont {ten
  Hagen}}, \bibinfo {author} {\bibfnamefont {R.}~\bibnamefont {Wittkowski}}, \
  and\ \bibinfo {author} {\bibfnamefont {H.}~\bibnamefont {L\"owen}},\
  }\href@noop {} {\bibfield  {journal} {\bibinfo  {journal} {Phys. Rev. E},\
  }\textbf {\bibinfo {volume} {84}},\ \bibinfo {pages} {031105} (\bibinfo
  {year} {2011}{\natexlab{b}})}\BibitemShut {NoStop}%
\bibitem [{\citenamefont {Wensink}\ and\ \citenamefont
  {L\"owen}(2008)}]{Wensink2008}%
  \BibitemOpen
  \bibfield  {author} {\bibinfo {author} {\bibfnamefont {H.~H.}\ \bibnamefont
  {Wensink}}\ and\ \bibinfo {author} {\bibfnamefont {H.}~\bibnamefont
  {L\"owen}},\ }\href@noop {} {\bibfield  {journal} {\bibinfo  {journal} {Phys.
  Rev. E},\ }\textbf {\bibinfo {volume} {78}},\ \bibinfo {pages} {031409}
  (\bibinfo {year} {2008})}\BibitemShut {NoStop}%
\bibitem [{\citenamefont {Wensink}\ and\ \citenamefont
  {L\"owen}(2012)}]{WensinkJPCM}%
  \BibitemOpen
  \bibfield  {author} {\bibinfo {author} {\bibfnamefont {H.~H.}\ \bibnamefont
  {Wensink}}\ and\ \bibinfo {author} {\bibfnamefont {H.}~\bibnamefont
  {L\"owen}},\ }\href@noop {} {\bibfield  {journal} {\bibinfo  {journal} {J.
  Phys.: Condens. Matter},\ }\textbf {\bibinfo {volume} {24}},\ \bibinfo
  {pages} {464130} (\bibinfo {year} {2012})}\BibitemShut {NoStop}%
\bibitem [{\citenamefont {Kaiser}\ and\ \citenamefont
  {L\"owen}(2013)}]{KaiserPRE}%
  \BibitemOpen
  \bibfield  {author} {\bibinfo {author} {\bibfnamefont {A.}~\bibnamefont
  {Kaiser}}\ and\ \bibinfo {author} {\bibfnamefont {H.}~\bibnamefont
  {L\"owen}},\ }\href@noop {} {\bibfield  {journal} {\bibinfo  {journal} {Phys.
  Rev. E},\ }\textbf {\bibinfo {volume} {87}},\ \bibinfo {pages} {032712}
  (\bibinfo {year} {2013})}\BibitemShut {NoStop}%
\bibitem [{\citenamefont {Lauga}\ and\ \citenamefont
  {Powers}(2009)}]{Lauga_review:09}%
  \BibitemOpen
  \bibfield  {author} {\bibinfo {author} {\bibfnamefont {E.}~\bibnamefont
  {Lauga}}\ and\ \bibinfo {author} {\bibfnamefont {T.~R.}\ \bibnamefont
  {Powers}},\ }\href@noop {} {\bibfield  {journal} {\bibinfo  {journal} {Rep.
  Prog. Phys.},\ }\textbf {\bibinfo {volume} {72}},\ \bibinfo {pages} {096601}
  (\bibinfo {year} {2009})}\BibitemShut {NoStop}%
\end{thebibliography}%

\end{document}